\documentclass[useAMS,usenatbib]{mn2e}

 \usepackage{times}
 \usepackage{graphicx}

\def\pcm3{{\rm\thinspace cm$^{-3}$}}

\def\contcaption{\@conttrue\SFB@caption\@captype}

\title[Near-IR spectroscopy of BD candidates in Upper Sco]
{Near-infrared cross-dispersed spectroscopy of brown dwarf candidates 
in the Upper Sco association
\thanks{Based on observations obtained with the Gemini Observatory.}}
\author[N. Lodieu et al.]{N. Lodieu$^{1}$\thanks{E-mail: nlodieu@iac.es},  
N. C. Hambly$^{2}$, R. F. Jameson$^{3}$, S. T. Hodgkin$^{4}$ \\
$^{1}$Instituto de Astrof\'isica de Canarias, V\'ia L\'actea s/n,
E-38205 La Laguna, Tenerife, Spain \\
$^{2}$Scottish Universities' Physics Alliance (SUPA),
Institute for Astronomy, School of Physics, University of Edinburgh,
Royal Observatory, Blackford Hill, \\
Edinburgh EH9 3HJ, UK \\
$^{3}$Department of Physics and Astronomy, University of Leicester, 
University Road, Leicester LE1 7RH, UK \\
$^{4}$Institute of Astronomy, Madingley Road, Cambridge CB3 0HA, UK}

\begin{document}

\date{Accepted \today{}. Received \today{}; in original form \today{}}

\pagerange{\pageref{firstpage}--\pageref{lastpage}} \pubyear{2007}

\maketitle

\label{firstpage}

\begin{abstract}
We present near-infrared (1.15--2.50 microns) medium-resolution 
(R = 1700) spectroscopy of a sample of 23 brown dwarf candidates 
in the young Upper Sco association. We confirm membership of 21 
brown dwarfs based on their spectral shape, comparison
with field dwarfs, and presence of weak gravity-sensitive features.
Their spectral types range from M8 to L2 with an uncertainty of
a subclass, suggesting effective temperatures between 2700 and 
1800 K with an uncertainty up to 300 K and masses in the 
30--8 M$_{\rm Jup}$ range.
Among the non-members, we have uncovered a field L2 dwarf
at a distance of 120--140 pc, assuming that it is single.
The success rate of our photometric selection based on five
photometric passbands and complemented partly by proper
motion is over 90\%, a very promising result for future studies
of the low-mass star and brown dwarf populations in young open 
clusters by the UKIDSS Galactic Cluster Survey.
We observe a large dispersion in the magnitude versus spectral-type
relation which is likely the result of the combination of
several effects including age dispersion, extent and depth
of the association, a high degree of multiplicity and the occurrence of disks.
\end{abstract}


\section{Introduction}

Since the discovery of the first genuine brown dwarfs
\citep{rebolo95,nakajima95}, several hundreds of brown dwarfs have
been found isolated in the field \citep[see review by][]{kirkpatrick05},
and as companions \citep[see review by][]{burgasser07a}. The number
of substellar objects unearthed in clusters has increased 
considerably due to the advent of larger detectors and more 
efficient instrumentation, allowing for an efficient photometric
selection of candidates and then efficient spectroscopic confirmation
of the presence of low-gravity features for
young (age $\leq$ 10 Myr) brown dwarfs in regions like
Orion \citep{hillenbrand97,muench02,slesnick04},
$\sigma$\,Orionis \citep{zapatero00,bejar01,caballero07d},
IC\,348 \citep{luhman03b,muench03}, 
Taurus \citep{briceno02,luhman04a,guieu06}, 
$\rho$\,Ophiuchus \citep{luhman99b,wilking05}, 
Chameleon \citep{comeron04,luhman04b}, and 
Upper Sco \citep{slesnick06,lodieu06,lodieu07a}.

The classification of old field L-type dwarfs, a class of objects 
cooler than M dwarfs with effective temperatures (T$_{\rm eff}$)
between $\sim$2200--1500 K \citep{basri00,leggett00} in the 
field, is now well established at optical wavelengths 
\citep{kirkpatrick99,martin99a}. An infrared classification,
based on the strength of absorption bands and atomic lines observed
at low resolution (R $\sim$ 50--600), was later set up to match the 
optical spectral types \citep{reid01a,testi01,geballe02}. 
In addition, \citet[][hereafter M03]{mclean03} produced a homogeneous 
set of intermediate resolution (R $\sim$ 2000) spectra obtained with
the Near-Infrared Spectrograph (NIRSPEC) on Keck II \citep{mclean98}
and developed a set of spectral indices to derive a classification
based on this higher resolution data. Likewise, \citet{cushing05}
generated a set of spectra at similar resolution but over a larger
wavelength range (0.6--4.1 $\mu$m).

However, the current spectral classification does not take into 
account two important parameters: gravity and metallicity 
\citep{kirkpatrick05}. Indeed, the number of spectroscopic L dwarfs
reported to date in young clusters and star forming regions
\citep{metchev06,allers07,mohanty07} is insufficient 
to develop a spectral 
classification for young brown dwarfs. The tools currently
available comprise several methods. Firstly, the spectral shape
and the changes observed in the strength of gravity-sensitive
features is often employed to assess the youth of brown dwarfs
when compared to known field dwarfs
\citep{lucas01a,gorlova03,slesnick04,mcgovern04,wilking04,meeus05,allers07}.
Secondly, \citet{luhman99a} proposed a T$_{\rm eff}$ vs spectral type 
relation for young dwarfs, that relation later being revised by 
\citet{luhman03b} and extended by \citet{allers07}.
Finally, the direct comparison of observed spectra with theoretical
models \citep{allard00} allows an estimate of T$_{\rm eff}$, gravity, 
and masses \citep{lucas01a,meeus05}.

The Upper Sco (hereafter USco) OB association is part of the Scorpius 
Centaurus complex \citep{deGeus89}, located at a distance of 145$\pm$2 pc 
from the Sun 
\citep{deBruijne97,deZeeuw99}. The age of USco is 5 Myr with little 
scatter \citep{preibisch99}. The region is relatively free of extinction 
(Av $\leq$ 2 mag) and star formation has already ended 
\citep{walter94}. The association has been targeted at multiple 
wavelengths over the past decade starting with X-rays 
\citep{walter94,preibisch98,kunkel99,preibisch99}
but also with Hipparcos \citep{deBruijne97} as well as in the optical 
\citep{preibisch01,preibisch02,ardila00,martin04,slesnick06} and 
in the near-infrared \citep{lodieu06,lodieu07a}. Several tens of 
low-mass stars and brown dwarfs with spectral types later than M6 
have been confirmed as genuine spectroscopic substellar members 
in USco \citep{ardila00,martin04,slesnick06,lodieu06}.

Recently, we have found a large number of new BD candidates 
with spectral types later than M6 and estimated masses below 
0.05 M$_{\odot}$  from several square degrees imaged by the 
UKIRT Infrared Deep Sky Survey \citep[UKIDSS;][]{lawrence07} 
Galactic Cluster Survey 
(GCS). Additionally, we have extended the mass function down to 
0.01 M$_{\odot}$ with proper motion confirmation for sources more 
massive than 0.02 M$_{\odot}$ \citep{lodieu06,lodieu07a}.
In this paper we present the near-infrared cross-dispersed spectra 
of the 23 faintest BDs in Upper Sco extracted from the UKIDSS GCS
as well as two known late-M dwarfs used as spectral templates
\citep{martin04,slesnick06}.
In Section \ref{USco:obs_dr}
we detail the observations and the data reduction of the cross-dispersed 
spectra obtained with the Gemini Near-InfraRed Spectrograph 
\citep[GNIRS;][]{elias06a}. In Section \ref{USco:memb}
we assess the membership of all sources using the spectral shape,
gravity-sensitive features, and direct comparison with near-infrared
spectra of field dwarfs at similar resolution.
In Section \ref{USco:discussion} we discuss the implications
of our spectroscopic follow-up and the current uncertainties
on the determination of the effective temperatures and masses
of these young brown dwarfs.

%
%
%
%

%
%
%
\begin{table*}
  \caption{Photometric candidates in Upper Sco with near-infrared
  spectra. We list the IAU name, coordinates (J2000), WFCAM $JHK$ 
  photometry, the number of exposures and the on-source integration
  times in seconds (numbers in brackets represent the number of
  frames effectively used during the data reduction process),
  the spectral type (SpT) and the membership (Y$\equiv$Member; 
  N$\equiv$Non-member). Typical uncertainties on the photometry
  is better than 0.05 mag in $J$, 0.02 mag in $H$, and 0.01 mag
  in $K$ \citep[full photometry is available in][]{lodieu07a}. 
  Proper motions are accurate to 10 mas/yr. 
  Two known members, also observed with GNIRS, 
  are added to this list: DENIS1611 \citep[M9;][]{martin04}
  and SCH1625 \citep[M8;][]{slesnick06}. We list the 2MASS 
  photometry for these two sources because they lie outside the
  current GCS coverage.
}
  \label{tab_USco:list_cand}
  \begin{tabular}{c c c c c c c c c c c}
  \hline
IAU Name                   &    R.A.\    &      dec      &  $J$  &   $H$   &  $K$   & $\mu_{\alpha}\cos\delta$ & $\mu_{\delta}$ & ExpT & SpT & Memb \cr
  \hline
USco J154722.82$-$213914.3 & 15:47:22.82 & $-$21:39:14.3 & 15.64 & 14.83 & 14.18 &  $-$13 &  $-$31 &  8 (8)$\times$150 & L0 & Y \cr  
USco J155419.99$-$213543.1 & 15:54:19.99 & $-$21:35:43.1 & 14.93 & 14.28 & 13.71 &  $-$14 &  $-$18 &  8 (4)$\times$150 & M8 & Y \cr  
USco J160603.75$-$221930.0 & 16:06:03.75 & $-$22:19:30.0 & 15.85 & 15.10 & 14.44 &    --- &    --- &  4 (4)$\times$300 & L2 & Y \cr  
USco J160606.29$-$233513.3 & 16:06:06.29 & $-$23:35:13.3 & 16.20 & 15.54 & 14.97 &    --- &    --- &  4 (4)$\times$300 & L0 & Y \cr  
USco J160648.18$-$223040.1 & 16:06:48.18 & $-$22:30:40.1 & 14.93 & 14.35 & 13.84 &  $-$23 &  $-$15 &  8 (8)$\times$150 & M8 & Y \cr  
USco J160714.79$-$232101.2 & 16:07:14.79 & $-$23:21:01.2 & 16.56 & 15.83 & 15.07 &    --- &    --- &  4 (4)$\times$300 & L0 & Y \cr  
USco J160723.82$-$221102.0 & 16:07:23.82 & $-$22:11:02.0 & 15.20 & 14.56 & 14.01 &  $-$11 &  $-$31 &  8 (8)$\times$150 & L1 & Y \cr  
USco J160727.82$-$223904.0 & 16:07:27.82 & $-$22:39:04.0 & 16.81 & 16.09 & 15.39 &    --- &    --- &  6 (6)$\times$300 & L1 & Y \cr  
USco J160737.99$-$224247.0 & 16:07:37.99 & $-$22:42:47.0 & 16.76 & 16.00 & 15.33 &    --- &    --- &  8 (8)$\times$300 & L0 & Y \cr  
USco J160818.43$-$223225.0 & 16:08:18.43 & $-$22:32:25.0 & 16.01 & 15.44 & 14.70 &    --- &    --- &  8 (8)$\times$300 & L0 & Y \cr  
USco J160828.47$-$231510.4 & 16:08:28.47 & $-$23:15:10.4 & 15.45 & 14.78 & 14.16 &  $-$12 &  $-$13 &  8 (4)$\times$150 & L1 & Y \cr  
USco J160830.49$-$233511.0 & 16:08:30.49 & $-$23:35:11.0 & 14.88 & 14.29 & 13.76 &  $-$5  &  $-$12 &  8 (8)$\times$150 & M9 & Y \cr  
USco J160847.44$-$223547.9 & 16:08:47.44 & $-$22:35:47.9 & 15.69 & 15.09 & 14.53 &     0  &  $-$20 &  4 (4)$\times$300 & M9 & Y \cr  
USco J160843.44$-$224516.0 & 16:08:43.44 & $-$22:45:16.0 & 18.58 & 17.22 & 16.26 &    --- &    --- & 12 (8)$\times$300 & L1 & Y \cr  
USco J160918.69$-$222923.7 & 16:09:18.69 & $-$22:29:23.7 & 18.08 & 17.06 & 16.16 &    --- &    --- &  8 (8)$\times$300 & L1 & Y \cr  
USco J160956.34$-$222245.5 & 16:09:56.34 & $-$22:22:45.5 & 17.83 & 16.99 & 16.29 &    --- &    --- &  8 (8)$\times$300 & dL2 & N \cr  
USco J161047.13$-$223949.4 & 16:10:47.13 & $-$22:39:49.4 & 15.26 & 14.57 & 14.01 &  $-$15 &  $-$24 &  8 (8)$\times$150 & M9 & Y \cr  
DENIS 161103.60$-$242642.9 & 16:11:03.60 & $-$24:26:42.9 & 14.86 & 14.14 & 13.70 &    --- &    --- &  8 (8)$\times$150 & M9 & Y \cr  
USco J161227.64$-$215640.8 & 16:12:27.64 & $-$21:56:40.8 & 17.14 & 16.37 & 15.78 &    --- &    --- &  8 (4)$\times$300 & L0 & ? \cr  
USco J161228.95$-$215936.1 & 16:12:28.95 & $-$21:59:36.1 & 16.41 & 15.56 & 14.79 &    --- &    --- &  4 (4)$\times$300 & L1 & Y \cr  
USco J161302.32$-$212428.4 & 16:13:02.32 & $-$21:24:28.4 & 17.17 & 16.37 & 15.65 &    --- &    --- &  8 (8)$\times$300 & L0 & Y \cr  
USco J161421.44$-$233914.8 & 16:14:21.44 & $-$23:39:14.8 & 14.97 & 14.44 & 13.94 &     0  &  $-$18 &  8 (8)$\times$150 & A0 & N \cr  
USco J161441.68$-$235105.9 & 16:14:41.68 & $-$23:51:05.9 & 16.07 & 15.34 & 14.62 &    --- &    --- &  4 (4)$\times$300 & L1 & Y \cr  
SCH 162528.62$-$165850.6   & 16:25:28.62 & $-$16:58:50.6 & 13.68 & 13.01 & 12.63 &    --- &    --- &  8 (8)$\times$150 & M8 & Y \cr  
USco J163919.15$-$253409.9 & 16:39:19.15 & $-$25:34:09.9 & 17.20 & 16.39 & 15.61 &  $-$1  &  $-$17 &  8 (8)$\times$300 & L1 & Y \cr  
 \hline
\end{tabular}
\end{table*}

%
%
\section{Observations and data reduction}
\label{USco:obs_dr}

In this section we describe the observations and data reduction of 
GNIRS spectra obtained for the faintest brown dwarf candidates 
extracted from two lists of photometric sources published in 
\citet{lodieu06} and \citep{lodieu07a}.
Their $H$ magnitudes are fainter than 14 mag and their estimated
masses below $\sim$25 M$_{\rm Jup}$, according to theoretical
models at 5 Myr \citep{baraffe98,chabrier00c}.
About half of them have proper motion measurements from the
cross-correlation between 2MASS and the GCS
(Table \ref{tab_USco:list_cand}).
The photometry with their associated errors and the proper motions
can be retrieved from \citet{lodieu06} and \citet{lodieu07a} but
we give again those data in Table \ref{tab_USco:list_cand}
for consistency.

%
%
\subsection{Observations}
\label{USco:obs}

Spectroscopic observations were conducted with the GNIRS spectrograph
\citep{elias06a} mounted on the Gemini South telescope in March
(09, 21, 24, 26, 31) and April (01, 02, 04, 06, 07) 2007 as part of 
the programme GS-2007A-Q-12 (Lodieu, PI). The list of USco targets
contains 23 photometric brown dwarf candidates fainter than $H$ = 14 mag
 and two known members, SCH162528.62$-$165850.55
\citep[hereafter SCH1625 (M8; $J_{\rm 2MASS}$ = 13.68);][]{slesnick06} 
and DENIS161103.6$-$242642.9 
\citep[hereafter DENIS1611 (M9; $J_{\rm DENIS}$ = 14.68);][]{martin04}.
GNIRS was used in cross-dispersed mode with the
32l/mm grism and a slit of 0.3 arcsec, yielding a coverage of 
1.15--2.5 microns at a resolution of R=1700 per resolution element. 
Only orders 3, 4, and 5 are usable due to severe inter-order 
contamination affecting the others. The total integration times
and the shift pattern were chosen as a function of the brightness 
of the objects (Table \ref{tab_USco:list_cand}): typically we have
on-source integration times of 150 or 300 seconds with an ABBA pattern
repeated one to three times. The number of exposures and the on-source
integration times are given in Table \ref{tab_USco:list_cand}.
The number in brackets represents the number of exposures effectively
used during the data reduction process.
Wavelength calibration was achieved using internal argon lamps with an 
rms better than 0.3\AA{}. The effects of telluric absorption were removed 
using two A0 standard stars (HIP79229 and HIP79244) observed immediately 
before or after the target and at a similar airmass (all objects are 
within an area of 2$\times$3 degrees in USco). The observations were 
conducted under the following conditions: image quality better than 
85\% (i.e.\ seeing between 0.6 and 0.8 arcsec), photometric 
conditions (cloud cover less than 50\% according to the Gemini
definition for proposals), and no restrictions
on the sky background and water vapour.

%
%
\subsection{Data reduction}
\label{USco:dr}

Data reduction was carried out following the instructions given in the 
IRAF files provided by Gemini Observatory to deal with cross-dispersed 
spectra taken with GNIRS\@. After correcting images for quadrant to
quadrant variations (task NVNOISE), we prepared the frames with
the task NSPREPARE in order to update
the headers and run the subsequent tasks. A final flat field image
was created using NSREDUCE and NSFLAT for each order. The telluric
standards and science targets were then trimmed, sky subtracted,
and flat-fielded using the same NSREDUCE command. Distortion correction
and wavelength calibration were achieved with the routines NSSDIST
and NSWAVELENGTH, respectively. Then, all the exposures of the
standard stars and science targets were combined into one
single file with NSCOMBINE before extracting a 1D spectrum with
NSEXTRACT\@. Note that some exposures were rejected in a few cases
due to the lack of signal or a high level of sky background
(ninth column of Table \ref{tab_USco:list_cand}).
We have interpolated linearly across the hydrogen 
recombination lines present in the spectrum of the A0 standard stars
\citep{vacca03} and clearly visible at our resolution (the $H$-band 
is the most affected). The interpolated lines are the Paschen $\gamma$ 
line at 12818\AA{} and Brackett lines at 15440\AA{}, 15560\AA{}, 
15708\AA{}, 15880\AA{}, 16110\AA{}, 16410\AA{}, 16810\AA{}, 
17370\AA{}, and 21655\AA{}.
Finally, we divided the 1D spectrum of the telluric 
standard from the 1D spectrum of the science target and multiplied 
by the spectrum of a A0 star available on the European Southern 
Observatory webpage\footnote{The spectra can be downloaded from the
URL: http://www.eso.org/ISAAC/}. 

\begin{figure}
   \centering
   \includegraphics[width=\linewidth]{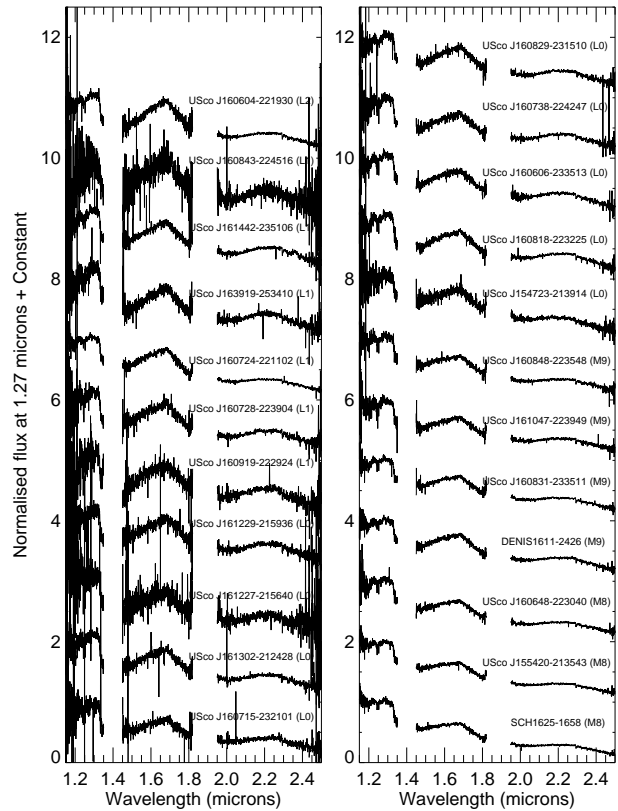}
   \caption{Cross-dispersed near-infrared (1.15--2.5 microns) spectra 
   of new USco brown dwarfs confirmed as spectroscopic members.
   Two known members are included for comparison: 
   SCH1625 \citep[M8;][]{slesnick06} 
   and DENIS1611 \citep[M9;][]{martin04}. The name of the object 
   and its associated spectral type are marked above the spectra.
   New members are ordered by increasing spectral types. Spectra
   are normalised at 1.27 microns and shifted along the y-axis for 
   clarity.
}
   \label{fig_USco:all_GNIRS_spec}
\end{figure}

The final near-infrared (1.15--2.50 microns) spectra of 21 photometric
brown dwarf candidates in USco confirmed spectroscopically as members 
are displayed in Fig.\ \ref{fig_USco:all_GNIRS_spec}.
The spectral region below 1.15 microns and the intervals 1.35--1.45
and 1.82--1.95 microns have been omitted due to the presence of
telluric absorption bands.
The location of the newly confirmed spectroscopic members in the
($J-K$,$K$) colour-magnitude diagram is shown in 
Fig.\ \ref{fig_USco:cmdJKK}.

%
%
%

%
%
\section{Classification and membership}
\label{USco:memb}

In this section we discuss the membership of 23 brown dwarf 
candidates in USco using the spectral shape of the near-infrared 
spectra, the comparison with known USco members and field dwarfs
classified optically, and the strength of gravity-sensitive features.

\subsection{Spectral types}
\label{USco:memb_SpT}

The spectral shape of our spectra indicate that 21 out of 23 photometric
candidates are indeed young brown dwarfs belonging to the USco
association. In particular, the triangular shape observed in the $H$-band
represents a characteristic feature of young objects 
\citep{lucas01a,meeus05}.
The youth of those brown dwarfs is further warranted by the presence
of weak Na {\small{I}} and K {\small{I}} doublets, spectral features
sensitive to gravity (Sect.\ \ref{USco:memb_grav}).
The GNIRS spectra of all new spectroscopically
confirmed members are shown in Fig.\ \ref{fig_USco:all_GNIRS_spec}.
Among the spectroscopic non-members we have one reddened
early-type star and one field L2 dwarf whose spectra matches well the
NIRSPEC spectrum of 2MASS J001544.8$+$351603
published by M03,\footnote{The infrared spectra from the NIRSPEC Brown
Dwarf Spectroscopic Survey \citep[BDSS;][]{mclean03} are available 
at the URL http://www.astro.ucla.edu/~mclean/BDSSarchive/}.

\begin{figure*}
   \centering
   \includegraphics[width=0.48\linewidth]{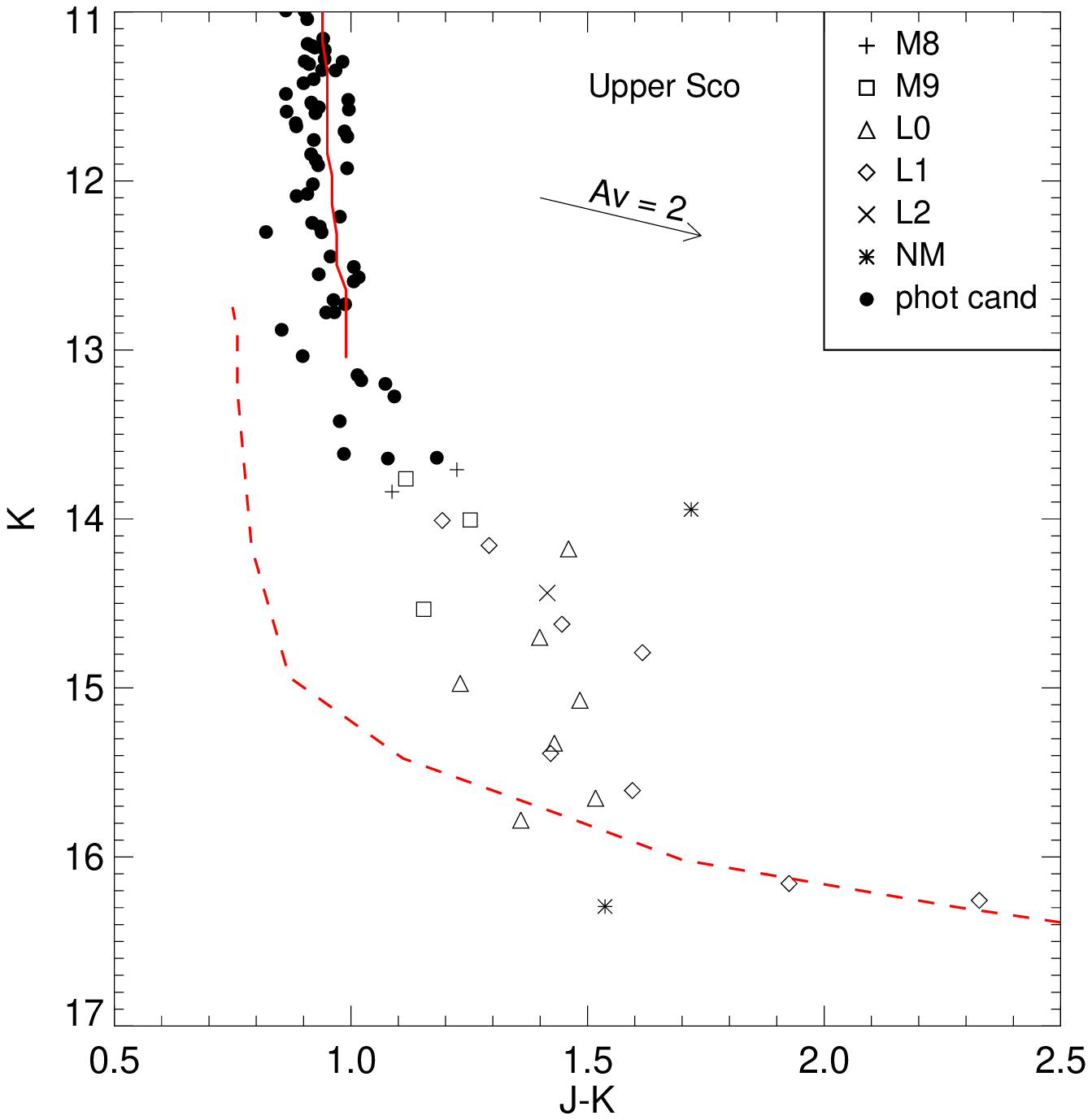}
   \includegraphics[width=0.48\linewidth]{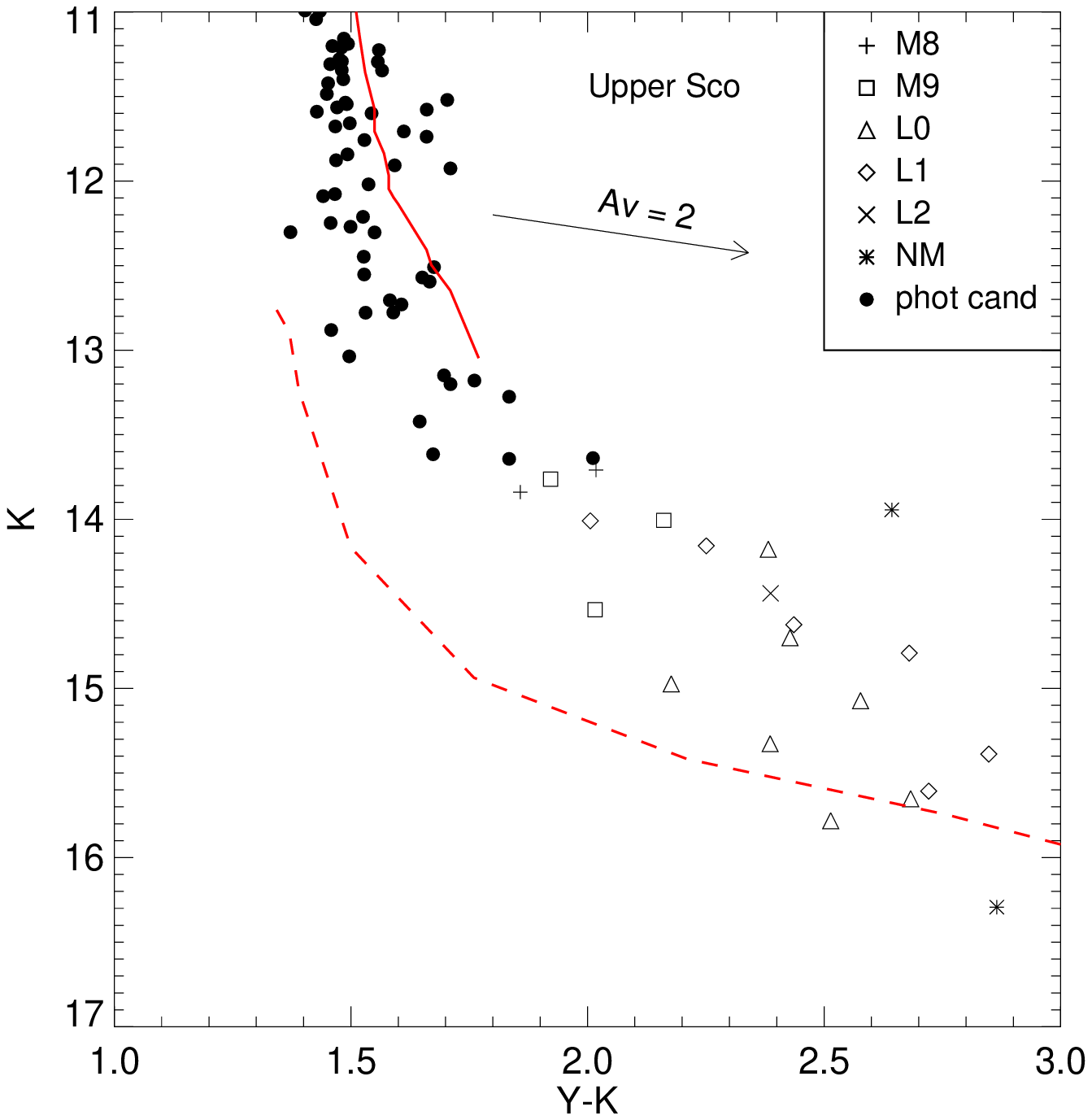}
   \caption{($J-K$,$K$) and ($Y-K$,$K$) colour-magnitude diagrams
for photometric candidates with no spectroscopy yet available
\citep[filled circles;][]{lodieu07a}, spectroscopic
non-members (asterisks) and newly confirmed spectroscopic members
in USco. Symbols represent the spectral types
assigned to the new members: pluses for M8, squares for M9,
triangles for L0, diamonds for L1, and crosses for L2 dwarfs.
Overplotted are the NextGen \citep[solid line;][]{baraffe98} and
DUSTY \citep[dashed line;][]{chabrier00c} isochrones for an age
of 5 Myr and shifted to a distance of 145 pc.
A reddening vector with A$_{\rm V}$ = 2 mag using the extinction
law from \citet{rieke85} has been added to illustrate the upper 
limit of the extinction observed in USco.
}
   \label{fig_USco:cmdJKK}
\end{figure*}

\begin{figure}
   \centering
   \includegraphics[width=0.99\linewidth]{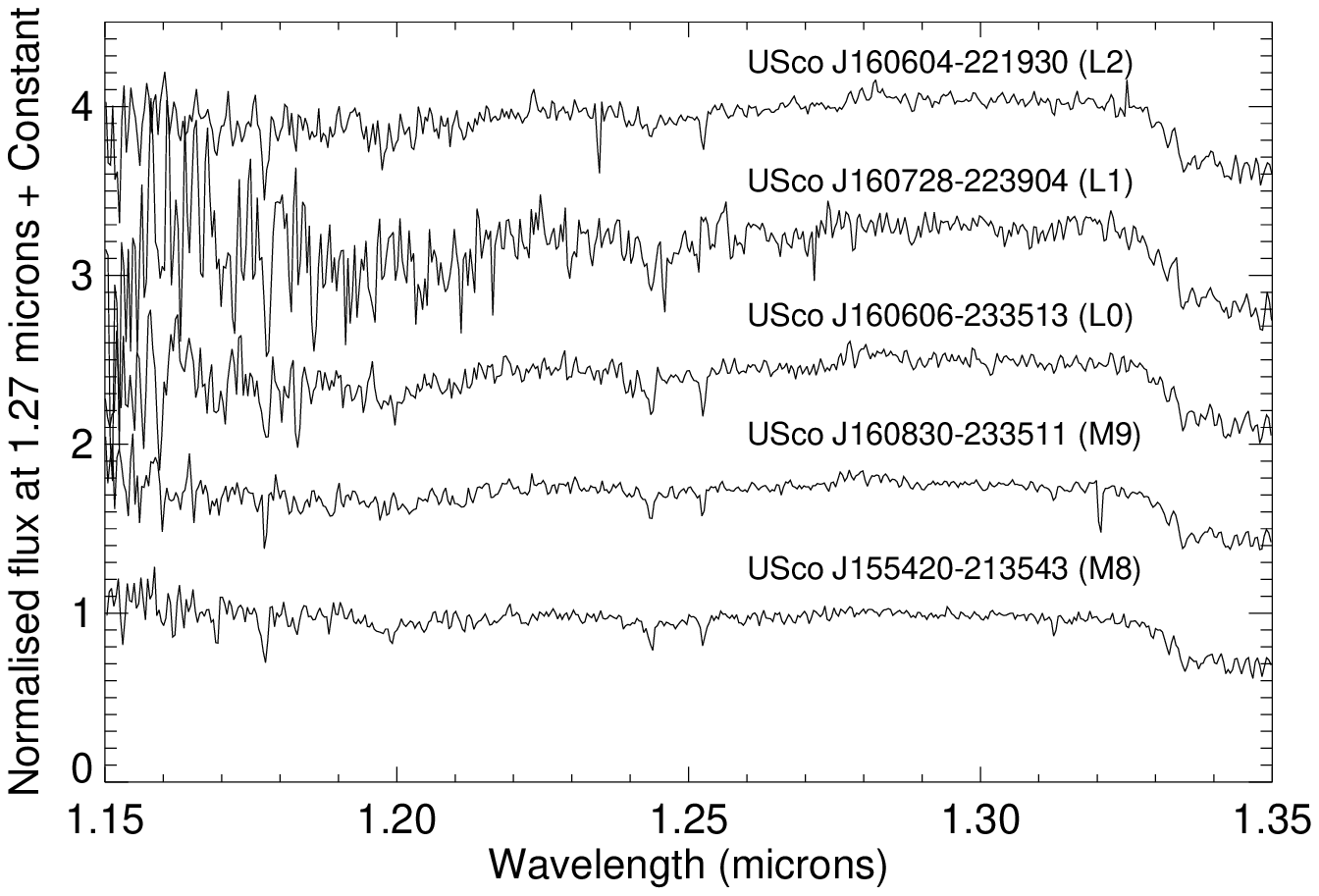}
   \includegraphics[width=0.99\linewidth]{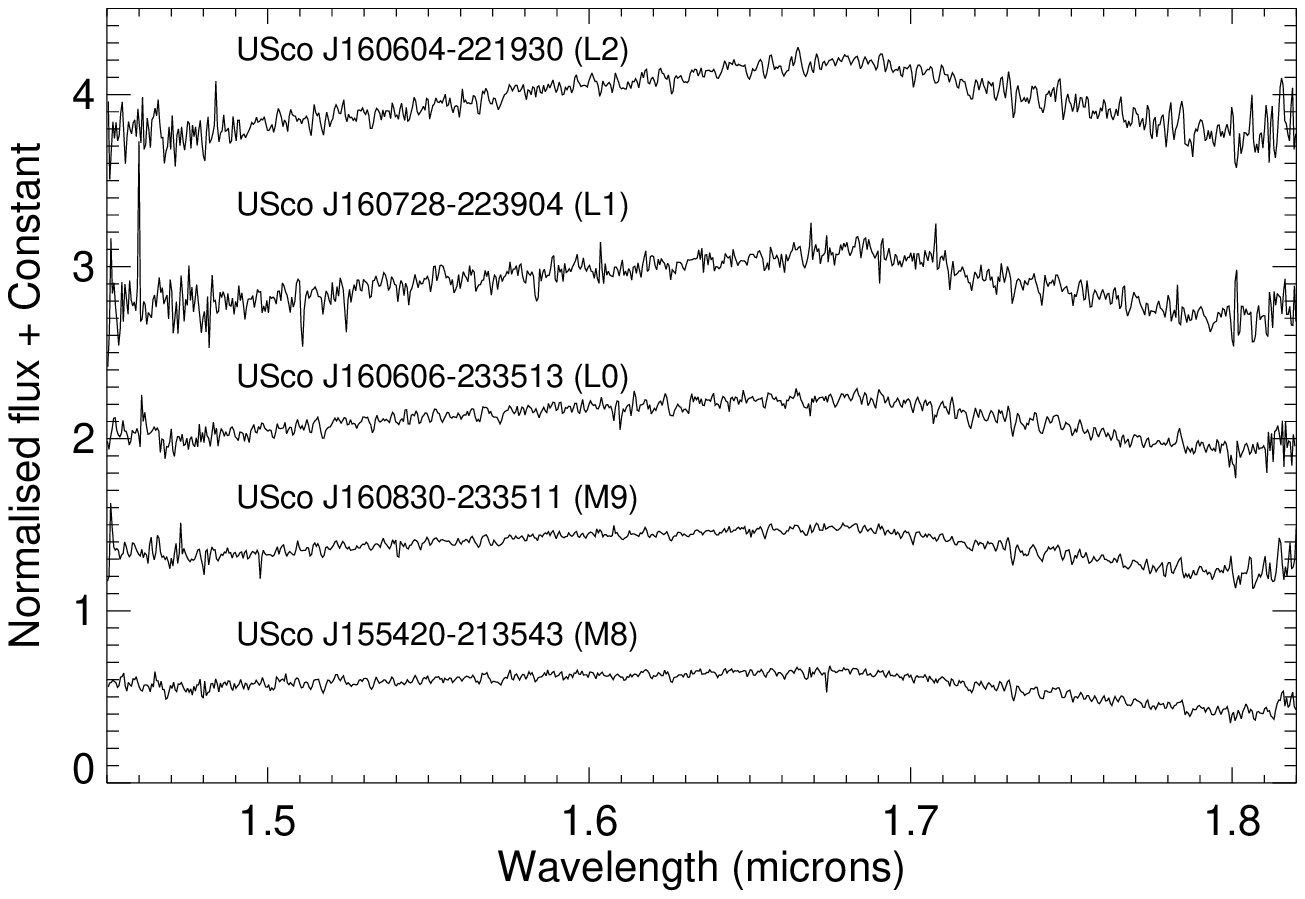}
   \includegraphics[width=0.99\linewidth]{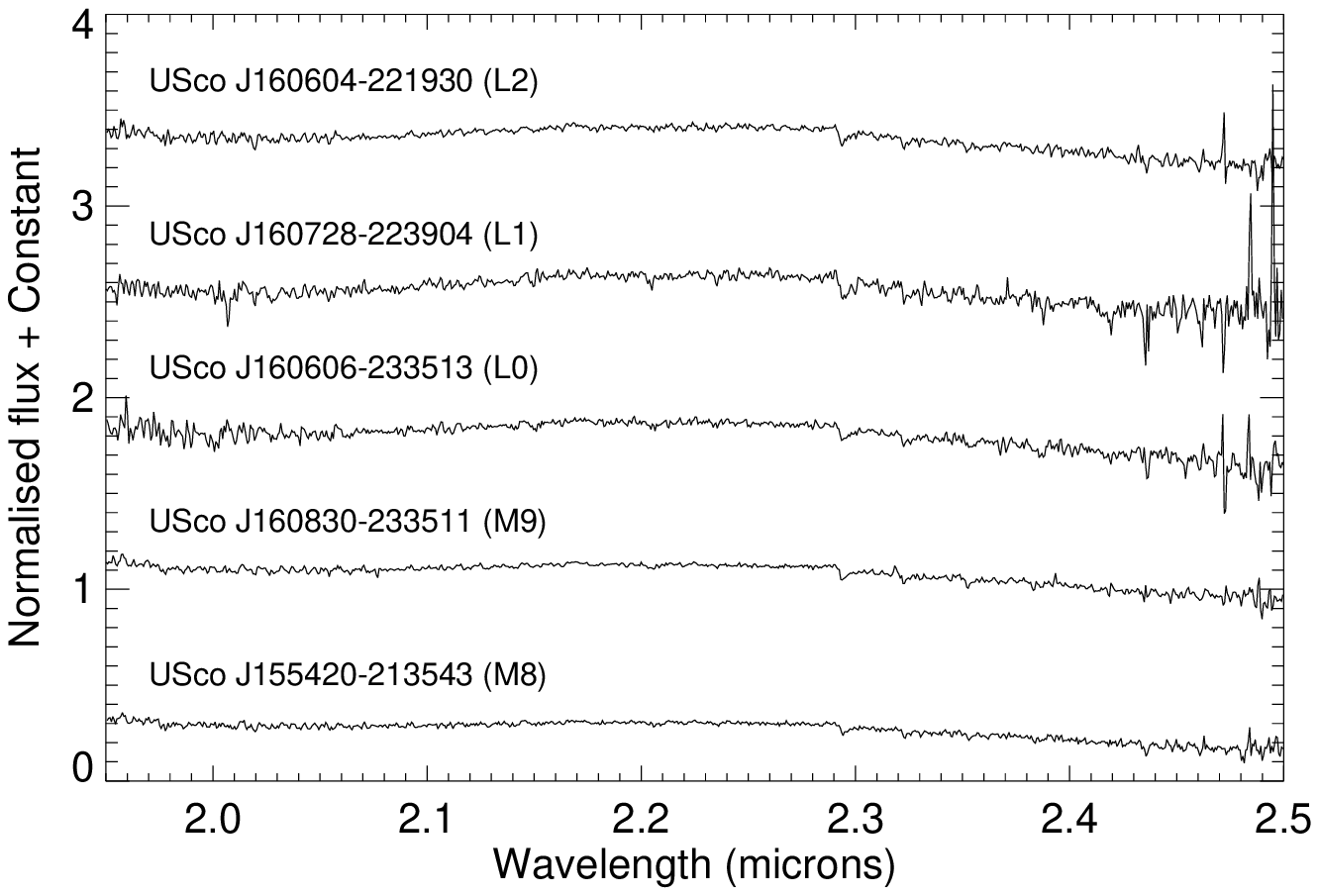}
   \caption{Close-ups on the $J$ (1.15--1.35 $\mu$m; top),
   $H$ (1.45--1.82 $\mu$m; middle), and $K$ (1.95--2.50 $\mu$m; bottom)
   portions of the GNIRS spectra of M8--L2 brown dwarfs in USco.
   From top to bottom are USco J155420$-$213543 (M8),
   USco J160830$-$233511 (M9), USco J160606$-$233513 (L0),
   USco J160728$-$223904 (L1), and USco J160604$-$221930 (L2).
   A constant of 0.75 was applied to offset the spectra.
}
   \label{fig_USco:MLsequence}
\end{figure}

We have compared the GNIRS spectra of all USco sources between each 
other to infer a spectral sequence. The sources selected as spectral
templates and used to classify the other targets are plotted in
Fig.\ \ref{fig_USco:MLsequence}. In addition, we have compared our
spectra with two known members classified optically: SCH1625
\citep[M8;][]{slesnick06} and DENIS1611 \citep[M9;][]{martin04}.
This procedure led to the four groups of objects to which we assign
spectral types spaced by a subtype (Fig.\ \ref{fig_USco:all_GNIRS_spec};
Tables \ref{tab_USco:list_cand} and \ref{tab_USco:list_indices}):
\begin{enumerate}
\item the first group contains two sources whose spectra match perfectly 
the infrared spectrum of SCH1625 (plusses in Fig.\ \ref{fig_USco:cmdJKK}). 
Hence we assign a spectral type of M8$\pm$0.5 following the 
optical classification by \citet{slesnick06};
\item the second group is made of three M9 dwarfs whose spectra agree
with the infrared spectrum of DENIS1611  (squares in 
Fig.\ \ref{fig_USco:cmdJKK}). The difference between the infrared 
spectra of the M8 and M9 dwarfs is characterised by (1) a steeper
$J$-band with later spectral type, (2) deeper H$_{2}$O band in the 
$H$-band with a peak shifted to a slightly redder wavelength, and
(3) deeper flux depletion longwards of 2.3 $\mu$m likely due to 
collision-induced absorption by molecular hydrogen. 
The depth of the CO break does not change significantly across 
the range of spectral types probed by our study;
\item the third group composed of eight sources exhibits a steeper 
$J$-band spectrum and a stronger depletion in the $K$-band beyond 
2.3 $\mu$m. The $H$-band is also affected with a strong water 
absorption in the 1.5--1.6 $\mu$m range. Hence, we classify those 
sources as L0 (triangles in Fig.\ \ref{fig_USco:cmdJKK}), a subclass 
later than DENIS1611. We assume an uncertainty of half a subclass 
on the spectral type;
\item the fourth group, made of seven sources, exhibit deeper H$_{2}$O 
absorption bands around 1.5 and 1.8 $\mu$m as well as a stronger
depletion in the $K$-band than sources in the previous group. 
The $J$-band seems however unaffected. Therefore, we assign 
a spectral type of L1$\pm$0.5 to all those sources
(diamonds in Fig.\ \ref{fig_USco:cmdJKK});
\item finally, one source, USco J160603.75$-$221930.0, exhibits even 
stronger water absorption in the $H$-band than the sources in the 
previous group although no difference is seen in the $J$- and $K$-bands. 
Thus we assign a spectral type of L2$\pm$1 to this source 
(cross in Fig.\ \ref{fig_USco:cmdJKK}), keeping in mind that the 
uncertainty in the classification might be larger than for the 
other sources. Nevertheless, direct comparison with medium-resolution
spectra of HD\,203030B \citep[L7.5$\pm$0.5;][]{metchev06} confirm 
that our latest object is of earlier type. Similarly, comparison
with the low-resolution (R$\sim$100) and low signal-to-noise (3--10)
spectrum of 2MASS1207$-$3932B \citep[mid- to late-L;][note that 
this object possess a disk]{mohanty07} suggest that 
USco J160603.75$-$221930.0 is earlier than a young mid-L dwarf.
%
\end{enumerate}

\begin{figure}
   \centering
   \includegraphics[width=\linewidth]{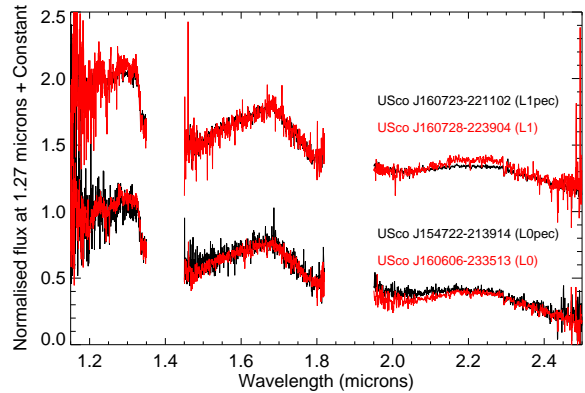}
   \caption{Spectra of two sources (black lines) showing spectral 
   shapes differing from our L0 (USco J160606$-$233513) and 
   L1 (USco J160728$-$223904) templates (red lines). 
   {\it{Bottom:}} ULAS J160723$-$221102 and
   {\it{top:}} ULAS J154722$-$213914\@.
}
   \label{fig_USco:peculiar}
\end{figure}

Finally, we have two sources with spectral shapes differing from the
groups described above: on the one hand, USco J154722.82$-$213914.3
is consistent with an L0 spectral type from $J$ to $K$ but exhibits 
an excess of flux in the 1.9--2.1 $\mu$m wavelength range which is 
difficult to explain (bottom in Fig.\ \ref{fig_USco:peculiar}). 
Furthermore, USco J160723.82$-$221102.0 appears warmer than an
L1 dwarf in the $J$-band but consistent with an L1 dwarf in the $H$-band. 
In addition, there is a clear excess of flux in the $K$-band at 
2.1--2.3 $\mu$m (top in Fig.\ \ref{fig_USco:peculiar}). 
This difference may result from a binary system composed of a L0 
and a L2 dwarf of roughly equivalent brightness. The $K$-band 
excess may be due to the presence of a disk although photometry at
longer wavelength is required to confirm this hypothesis. Another
option is that the system could present a temperature inversion
as the one observed in the eclipsing binary discovered in Orion by
\citet{stassun06}. Nonetheless, we tentatively assign a 
spectral type of L1 to that object.

%
%
%
%

%
%
%

\subsection{Spectral indices}
\label{USco:memb_index}

To assess the validity of our spectral classification, we have computed 
several spectral indices designed in the literature for young objects
(Table \ref{tab_USco:list_indices}), including three vapour indices
\citep[H$_{2}$O and H$_{2}$O-1+H$_{2}$O-2;][]{allers07,slesnick04}
as well as an FeH index \citep{slesnick04} and a K2 index 
\citep{tokunaga99}. Those indices were
originally defined to classify field L dwarfs in the infrared 
\citep{tokunaga99,mclean00a,reid01a,mclean03} matching their 
optical spectral classifications \citep{kirkpatrick99,martin99a}.
Subsequently several authors modified those indices to apply them
to younger brown dwarfs \citep{wilking99,gorlova03,slesnick04,allers07}.

The H$_{2}$O index ($<$1.550--1.560$>$/$<$1.492--1.502$>$) defined by 
\citet{allers07} allows an estimate of the
spectral type with an uncertainty of one subtype; this relation is 
valid from M5 to L0\@. The inferred spectral types agree with our 
classifications for most objects within the quoted 
uncertainties (Table \ref{tab_USco:list_indices}).
The exceptions are USco J160723.82$-$221102.0 and 
USco J154722.82$-$213914.3 whose 
spectral shapes are discussed above (Sect.\ \ref{USco:memb_SpT}) as 
well as USco J160604$-$221930 that we classified as L2 and outside
the range of validity of this spectral index.

Two additional water vapour indices were defined by \citet{slesnick04}
in the $J$ and $K$-bands following the work on field dwarfs by
\citet{mclean00a} and \citet{reid01a}. Index H$_{2}$O-1 is defined
by the ratio 1.34/1.30 microns whereas H$_{2}$O-2 is defined by the 
ratio 2.04/2.15 microns. Spectral types can be derived
from empirical relations given in Table 3 and shown in Fig.\ 5 of
\citet{slesnick04}. The inferred spectral types from both indices 
are consistent with our classification within the 1.5 subtype errors 
assigned to the empirical fit. The best agreement with our 
classification is obtained for the H$_{2}$O-2 index in the $K$-band.

\begin{figure}
   \includegraphics[width=0.99\linewidth]{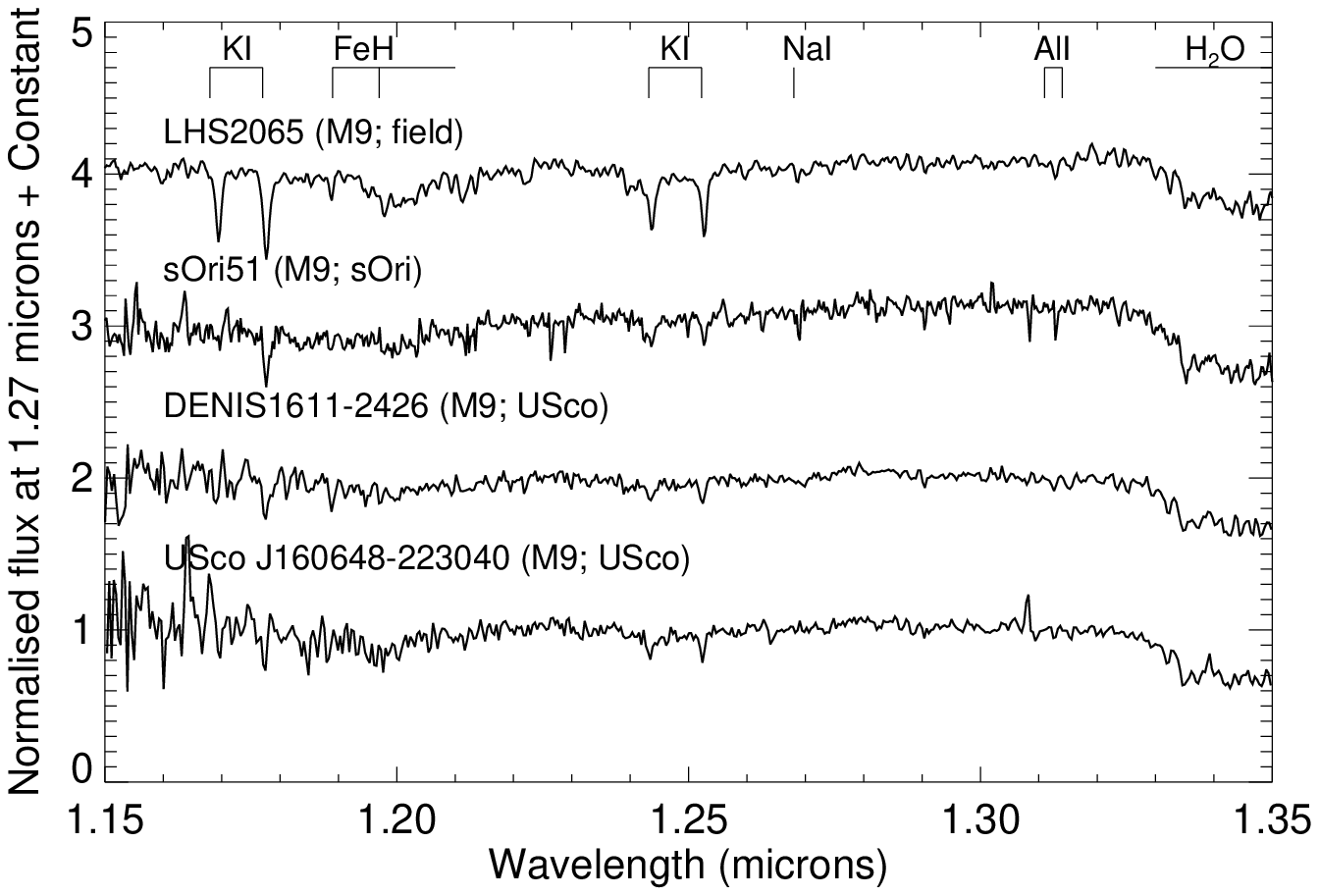}
   \includegraphics[width=0.99\linewidth]{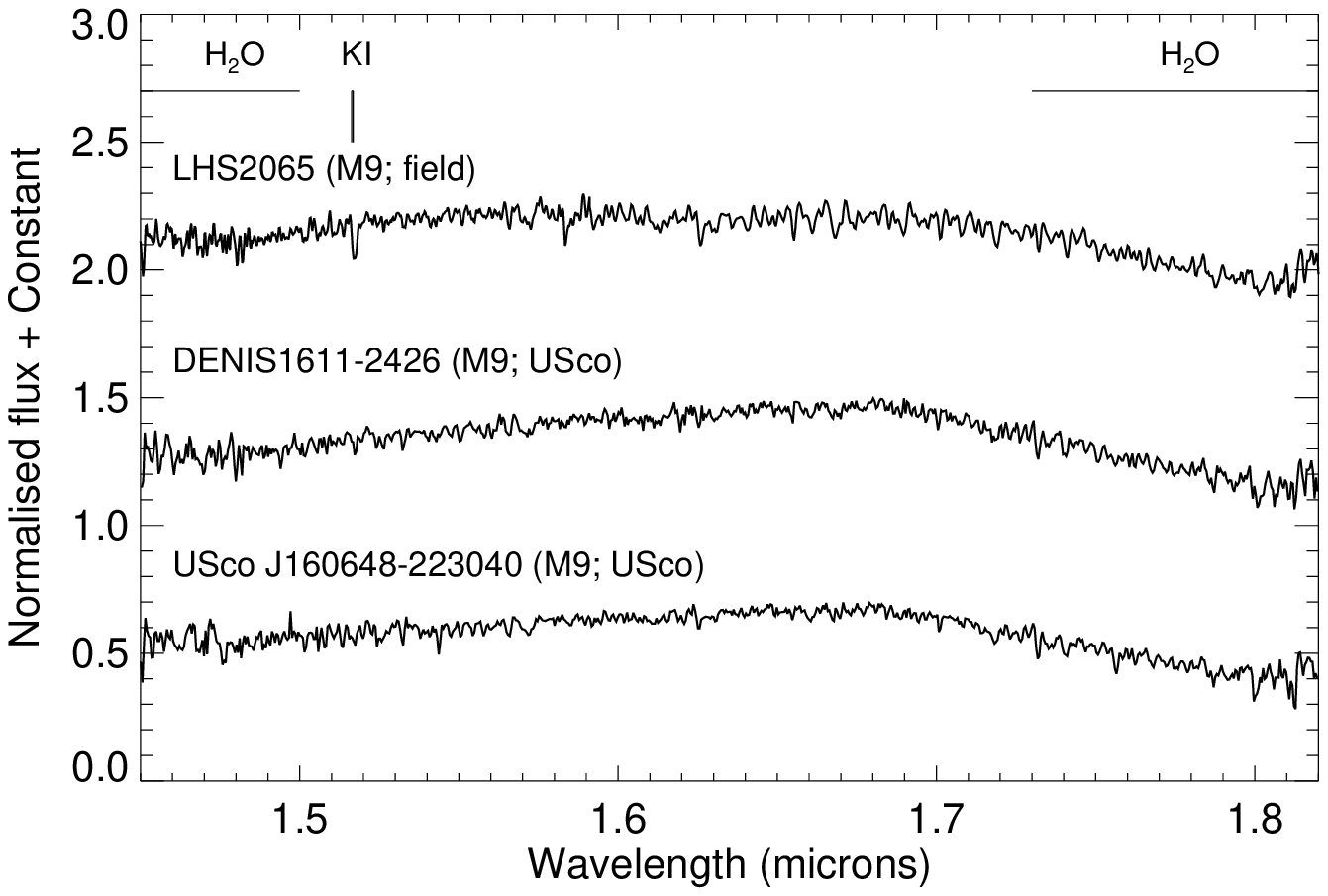}
   \includegraphics[width=0.99\linewidth]{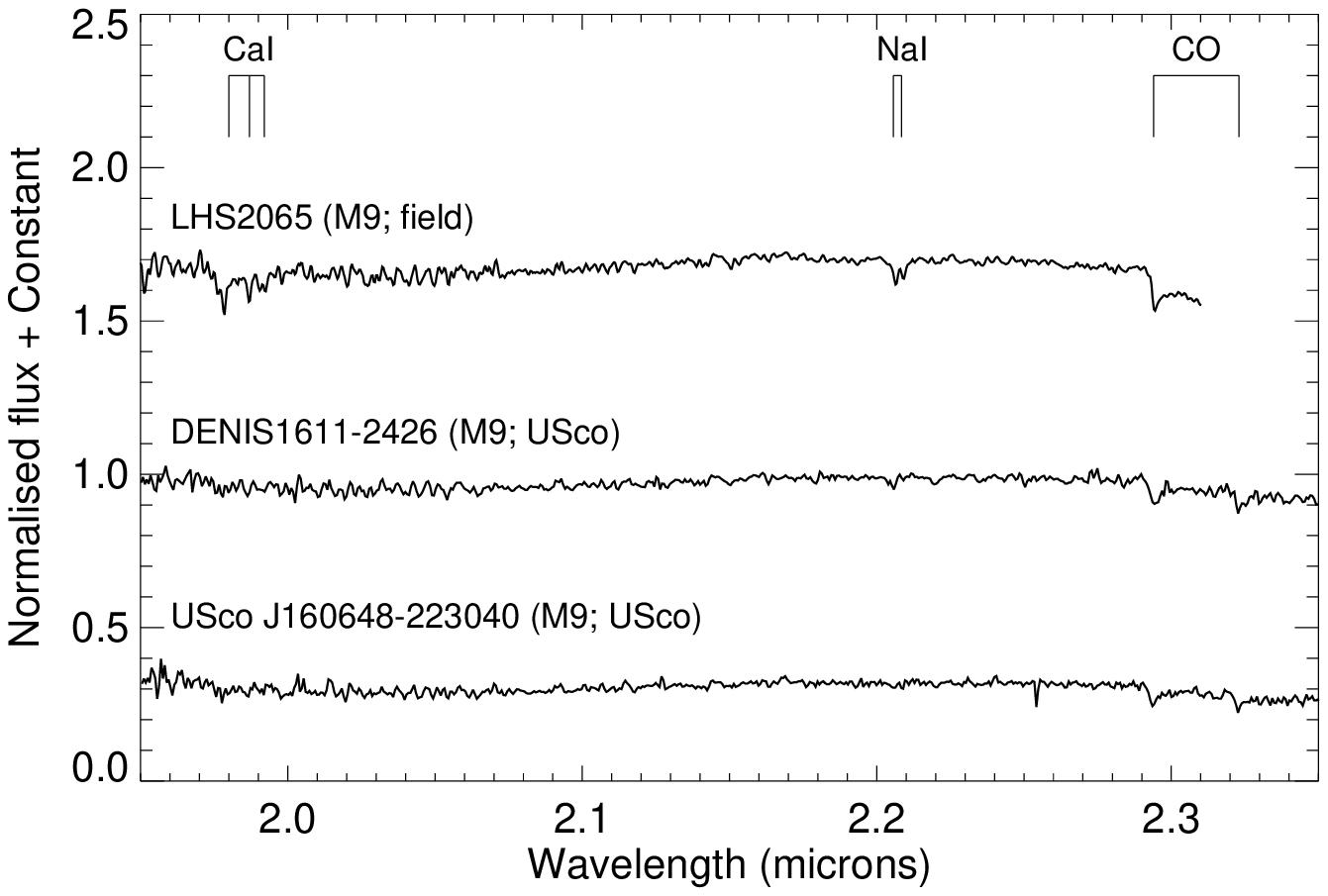}
   \caption{Near-infrared spectra of M9 dwarfs in $J$ (top),
   $H$ (middle), and $K$ (bottom panel) passbands, corresponding
   to three orders of the GNIRS instrument.
   We compare a new young M9 brown dwarf member of USco
   with DENIS1611, classified optically as M9 by
   \citet{martin04}, and a field M9 dwarf, LHS2065 whose
   spectrum was obtained with NIRSPEC on Keck II and published
   by M03.
   In the top panel, we have also added the spectrum of an M9
   dwarf belonging to the $\sigma$ Orionis cluster, sOri 51
   \citep{zapatero00,barrado01c} and observed with NIRSPEC
   \citep{mcgovern04}. The weak spectral lines present in
   the spectra of the USco member confirms a low gravity 
   and hence a young age for this object. 
   Spectra are shifted by a constant of unity for clarity.
}
   \label{fig_USco:M9compare}
\end{figure}

In addition, \citet{slesnick04} created a new index based on the
strength of the FeH band around 1.2 $\mu$m. This band is weak in our
spectra, suggesting that this feature depends on gravity
(Fig.\ \ref{fig_USco:M9compare}; Sect.\ \ref{USco:memb_grav}). Moreover,
our spectra are usually noisy in this region. The inferred indices
range from 0.8 to 1.0 (Table \ref{tab_USco:list_indices}), corresponding
to spectral types between M5 to L0 but, in general, the derived
spectral types differ significantly from our classification.
This discrepancy is confirmed by the spectral types assigned to
the optically classified young brown dwarfs DENIS1611 (M6 instead of M9)
and SCH1625 (M6 instead of M8). The origin of this inconsistency is 
unclear but may result from a combination of the large dispersion
observed in the empirical fit \citep[Table 3 and Fig.\ 5 in][]{slesnick04}
and the noisy GNIRS spectra in that wavelength range.

Furthermore, we have computed the H$_{2}$O 1.5 $\mu$m index defined
by the ratio of the flux between 1.57 and 1.59 $\mu$m and the
flux in the 1.46--1.48 $\mu$m range \citep{geballe02}. This
index probes the slope of the blue side of the $H$-band spectrum
and is defined for field L and T dwarfs. Hence, the inferred
spectral types might not be accurate but the values quoted in
Table \ref{tab_USco:list_indices} confirm the range in spectral
type (M8--L2) explored by our study.

Finally, we have applied the K2 index measuring the strength of
the H$_{2}$ absorption around 2.2 $\mu$m and proposed by 
\citet[][their equation 1]{tokunaga99}.
We have found positive values ranging from 0.0 to 0.05
(Table \ref{tab_USco:list_indices}) whereas typical values derived 
for field dwarfs are negative \citep[Fig.\ 4 in][]{tokunaga99},
suggesting a possible dependence on gravity of this index for 
late-M and early-L dwarfs.

%
%
%

%
%
%

\subsection{Comparison with field dwarfs}
\label{USco:memb_field}

A further test of our spectral classification is a direct comparison 
with field L dwarfs of similar spectral types. M03 published 
a sequence of near-infrared spectra of M, L, and T dwarfs taken at a 
comparable resolution as our spectra (R=2000 vs 1700) with NIRSPEC on 
Keck II. The spectra of vB\,10 (M8), LHS\,2065 (M9), 
2MASS J034543.2$+$254023 (L0), 2MASS J103524.6$+$250745 (L1), 
and 2MASS J001544.8$+$351603 \citep[L2;][]{kirkpatrick00}
observed over the full 1.15--2.30 $\mu$m range were downloaded from
the BDSS webpage. Those spectra of optically classified field dwarfs
(red lines) are plotted on top of our GNIRS spectra in 
Fig.\ \ref{fig_USco:youngVSold}. Close-ups on the $J$, $H$, and $K$
passbands are displayed in Fig.\ \ref{fig_USco:M9compare}
where important atomic lines and absorption bands are marked.

The agreement between the old field dwarfs and the young USco members
suggests that our spectral classification is quite reasonable. 
Starting with the M8 and M9 USco dwarfs for which optical classification 
is available from the literature \citep{martin04,slesnick06}, the main 
difference resides in the triangular shape of the $H$-band 
\citep{lucas01a,luhman04a,slesnick04,meeus05,allers07}. Additionally, 
the spectra of young objects appears slightly flatter around 1.3 $\mu$m.
The triangular shape of young sources remains to later spectral types, 
supporting a young age for our targets. Furthermore, the $H$- and $K$-band 
fluxes tend to be lower than those of field dwarfs, confirming the 
model trends of younger objects being bluer in $J-K$ than older 
objects \citep{chabrier00c}. Finally, an additional major difference is 
the weakness of gravity-sensitive features such as K {\small{I}} and 
Na {\small{I}} in our spectra, again implying low gravities and hence
supporting a young age for those 
brown dwarfs (see Sect.\ \ref{USco:memb_grav} for a
discussion on the strength of atomic lines).

\subsection{Gravity}
\label{USco:memb_grav}

Several atomic lines present in the near-infrared are known to be 
sensitive to gravity i.e.\ weaker with lower gravity or younger ages. 
We discuss in this section the behaviour of several of those lines 
seen in old late-M and early-L dwarfs (M03) and provide equivalent 
widths (EWs) in Table \ref{tab_USco:list_atomic}. Our measurements, 
in particular those
of the K {\small{I}} (1.168/1.177 $\mu$m and 1.243/1.254 $\mu$m), 
Na {\small{I}} (1.268 $\mu$m and 2.206/2.209 $\mu$m), and 
Ca {\small{I}} (1.98 $\mu$m) doublets, show that those 
gravity-sensitive features are weak in our spectra, hence 
adding a further proof to the youth of our targets.
However, we do not observe any obvious trend between the
strength of gravity-sensitive features (Table \ref{tab_USco:list_atomic})
and the position of the sources in the colour-magnitude diagrams
(Fig.\ \ref{fig_USco:cmdJKK}). Similarly, the current data do not
show a correlation between the position of the objects in the
colour-magnitude diagrams and their location in a ($l$,$b$)
plot \citep[Fig.\ 9 in][]{deZeeuw99}. Our photometric survey targeted 
the central part of the USco association with the highest concentration 
of members although marginal contamination by other subgroups (Upper
Centaurus Lupus and Lower Centaurus Crux) cannot be ruled out.

The K {\small{I}} doublets at 1.168/1.177 $\mu$m, 1.243/1.254 $\mu$m, 
and 1.517 $\mu$m are known to be sensitive to gravity. Indeed, all
the EWs given in Table \ref{tab_USco:list_atomic} are typically a 
factor of two smaller than those measured for field dwarfs of
similar spectral type \citep[M03;][]{cushing05}.
Na {\small{I}} is similarly sensitive to gravity and often 
used to disentangle young dwarfs in clusters from their old field 
counterparts at optical 
\citep{steele95a,martin96,zapatero97c,luhman99a,luhman04a,lodieu05a}
and infrared \citep{mcgovern04,allers07} wavelengths.
The Na {\small{I}} line at 1.268 $\mu$m is not detected at our resolution
and signal-to-noise whereas it is clearly present in the spectra of field 
late-M and early-L dwarfs \citep[M03;][]{cushing05}. However, this 
feature also depends on effective temperature because it disappears by 
mid-L but we are confident that the weakness of the line is due to 
gravity and not to effective temperature. We observe a similar behaviour 
for the Na {\small{I}} doublet at 2.206/2.209 $\mu$m with EWs typically 
lower than those of field dwarfs (Table \ref{tab_USco:list_atomic}).

\begin{figure*}
   \centering
   \includegraphics[width=\linewidth]{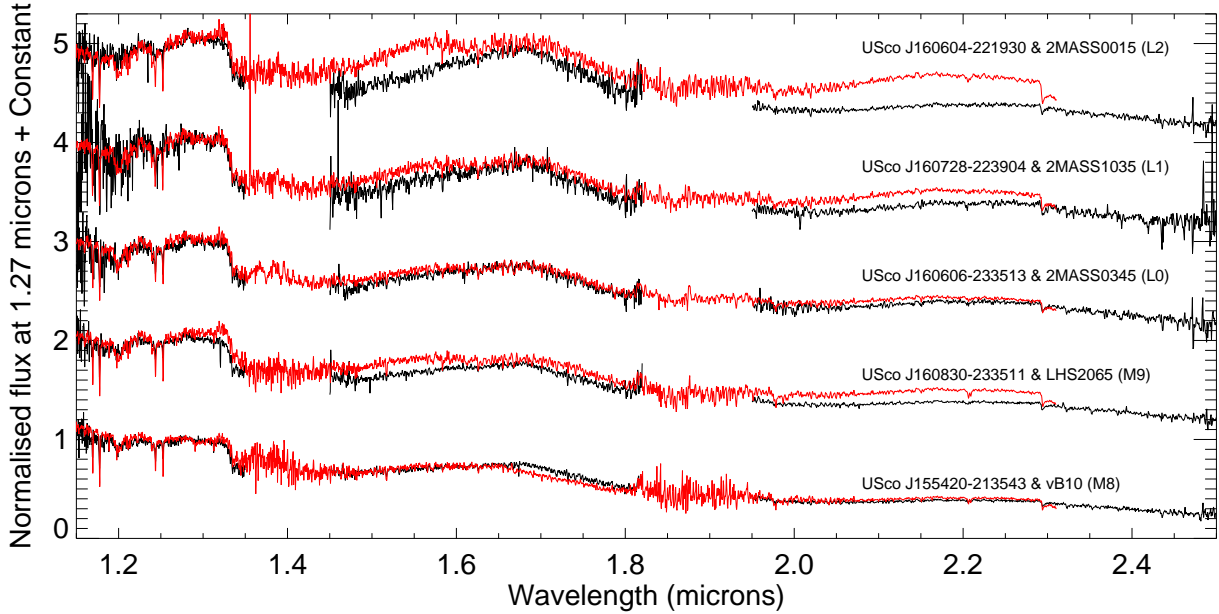}
   \caption{Near-infrared spectra of new
   young brown dwarfs in Upper Sco (black lines) compared to old
   field dwarfs \citep[red line;][]{mclean03} of similar spectral type.
   From top to bottom are USco J155420$-$213543 (M8),
   USco J160830$-$233511 (M9), USco J160606$-$233513 (L0),
   USco J160728$-$223904 (L1), and USco J160604$-$221930 (L2)
   compared with vB\,10 (M8), LHS\,2065 (M9), 2MASS0345$+$25 (L0),
   2MASS1035$+$35 (L1), and 2MASS0015$+$25 (L2), respectively.
   A constant of unity was applied to the spectra for clarity.
}
   \label{fig_USco:youngVSold}
\end{figure*}

The CO bandhead at 2.3 $\mu$m appears slightly weaker in our spectra
than those of field dwarfs (Fig.\ \ref{fig_USco:youngVSold}), suggesting
that it is sensitive to gravity as pointed out by M03. 
We do not find a strong dependence with spectral type and thus with 
effective temperature, again consistent with observations of field 
dwarfs (M03).
Furthermore, we do not detect the Ca {\small{I}} triplet at 1.98 $\mu$m
in our spectra whereas it is clearly detected in field late-M to 
early-L dwarfs before it disappears at cooler temperatures. Hence, 
this feature is strongly dependent on effective temperature 
\citep[M03;][]{cushing05} but also on gravity.

The Fe {\small{I}} doublet at 1.189/1.197 $\mu$m seems independent of
gravity because we have measured EWs (Table \ref{tab_USco:list_atomic})
comparable to those reported in Table 8 of M03 and in Table 9 of
\citet{cushing05}. However, the strength of that line is weak 
at all ages (EWs on the order of 1\AA{} or less) and higher 
resolution may subsequently reveal a dependence on gravity.
This could be the case since the shape of the spectra of young
brown dwarfs differs around 1.6 $\mu$m, a region affected by
FeH in older L dwarfs (Fig.\ \ref{fig_USco:youngVSold} \& 
\ref{fig_USco:M9compare}).

Finally, we have measured EWs for the Al {\small{I}} doublet at 
1.31/1.314 $\mu$m (the doublet is unresolved at our resolution) smaller 
for field dwarfs, suggesting that this feature is gravity dependent 
(Table \ref{tab_USco:list_atomic}; Table 8 in M03 and Table 9
in \cite{cushing05}).

%
%
\begin{table*}
  \caption{Spectral indices and their associated spectral types (Sp)
  for 21 brown dwarfs confirmed spectroscopically as members
  of USco. The L2 field dwarf classified as a non-member, 
  USco J160956.34$-$222245.5, is also included. We list water 
  vapour indices designed for young dwarfs: H$_{2}$O \citep{allers07},
  H$_{2}$O-1 and H$_{2}$O \citep{slesnick04},
  as well as the FeH \citep{slesnick04} and K2 \citep{tokunaga99}
  indices. We have also added in the last column the spectral index
  H$_{2}$O 1.5$\mu$m and its associated spectral type, defined for 
  field dwarfs by \citet{geballe02}.The second column lists the 
  adopted spectral type (Sp) from the direct comparison with previous 
  USco members (Table \ref{tab_USco:list_cand}). Objects are ordered 
  by increasing spectral type.
}
  \label{tab_USco:list_indices}
 \begin{tabular}{c c c c c c c l}
\hline
IAU Name              & Sp & H$_{2}$O (Sp) & H$_{2}$O-1 (Sp) & H$_{2}$O-2 (Sp) &  FeH (Sp) &  K2 & H$_{2}$O 1.5$\mu$m (Sp) \cr
\hline
SCH162528.62$-$165850.55   & M8 & 1.06 ( 7.02) &  0.70 ( 8.96) &  0.96 ( 8.06) &  0.95 ( 6.33) & -0.01 & 1.10 (M8) \cr 
USco J160648.18$-$223040.1 & M8 & 1.06 ( 7.10) &  0.68 ( 9.52) &  0.91 ( 9.34) &  0.85 ( 9.41) &  0.00 & 1.22 (L0) \cr 
USco J155419.99$-$213543.1 & M8 & 1.06 ( 7.01) &  0.71 ( 8.74) &  0.94 ( 8.58) &  0.94 ( 6.76) &  0.00 & 1.11 (M8) \cr 
DENIS161103.6$-$242642.9   & M9 & 1.13 ( 8.68) &  0.72 ( 8.21) &  0.92 ( 9.20) &  0.97 ( 5.79) &  0.02 & 1.23 (L0) \cr 
USco J161047.13$-$223949.4 & M9 & 1.12 ( 8.32) &  0.69 ( 9.48) &  0.90 ( 9.80) &  0.90 ( 7.91) &  0.03 & 1.20 (M9--L0) \cr 
USco J160830.49$-$233511.0 & M9 & 1.14 ( 8.90) &  0.69 ( 9.45) &  0.93 ( 8.84) &  0.90 ( 7.89) & -0.01 & 1.20 (M9--L0) \cr 
USco J160847.44$-$223547.9 & M9 & 1.12 ( 8.45) &  0.69 ( 9.43) &  0.93 ( 8.95) &  0.89 ( 8.16) &  0.02 & 1.23 (L0) \cr 
USco J154722.82$-$213914.3 & L0 & 1.07 ( 7.28) &  0.72 ( 8.19) &  0.98 ( 7.45) &  0.91 ( 7.55) & -0.01 & 1.10 (M8) \cr 
USco J160606.29$-$233513.3 & L0 & 1.15 ( 9.08) &  0.66 (10.51) &  0.88 (10.26) &  0.86 ( 9.25) &  0.03 & 1.27 (L0--L1) \cr 
USco J160714.79$-$232101.2 & L0 & 1.13 ( 8.74) &  0.73 ( 7.98) &  0.88 (10.28) &  0.91 ( 7.60) &  0.06 & 0.93 ($<$M8) \cr 
USco J160737.99$-$224247.0 & L0 & 1.17 ( 9.56) &  0.61 (12.24) &  0.82 (11.81) &  0.84 ( 9.86) &  0.04 & 1.29 (L1) \cr 
USco J160818.43$-$223225.0 & L0 & 1.21 (10.37) &  0.62 (11.97) &  0.88 (10.23) &  0.83 (10.17) &  0.03 & 1.32 (L1) \cr 
USco J160828.47$-$231510.4 & L0 & 1.08 ( 7.60) &  0.69 ( 9.46) &  0.88 (10.19) &  0.87 ( 8.95) &  0.03 & 1.26 (L0) \cr 
USco J160918.69$-$222923.7 & L1 & 1.28 (12.13) &  0.72 ( 8.34) &  0.83 (11.55) &  0.45 (22.12) &  0.04 & 1.55 (L4--L5) \cr 
USco J161227.64$-$215640.8 & L0 & 1.18 ( 9.73) &  0.67 (10.07) &  0.88 (10.36) &  0.97 ( 5.80) &  0.03 & 0.83 ($<$M8) \cr 
USco J161302.32$-$212428.2 & L0 & 1.22 (10.64) &  0.59 (12.71) &  0.87 (10.61) &  0.82 (10.54) &  0.03 & 1.24 (L0) \cr 
USco J160723.82$-$221102.0 & L1 & 1.13 ( 8.65) &  0.67 (10.02) &  0.88 (10.39) &  0.91 ( 7.47) &  0.02 & 1.23 (L0) \cr 
USco J160727.82$-$223904.0 & L1 & 1.25 (11.28) &  0.61 (12.22) &  0.83 (11.76) &  0.89 ( 8.31) &  0.04 & 1.26 (L0) \cr 
USco J160843.44$-$224516.0 & L1 & 1.22 (10.65) &  0.61 (12.13) &  0.80 (12.49) &  0.68 (14.91) &  0.01 & 1.53 (L4) \cr 
USco J161228.95$-$215936.1 & L1 & 1.18 ( 9.66) &  0.64 (11.14) &  0.87 (10.48) &  0.84 ( 9.91) &  0.03 & 0.83 ($<$M8) \cr 
USco J161441.68$-$235105.9 & L1 & 1.17 ( 9.49) &  0.61 (12.06) &  0.85 (10.97) &  0.86 ( 9.07) &  0.05 & 1.25 (L0) \cr 
USco J163919.15$-$253409.9 & L1 & 1.14 ( 8.88) &  0.63 (11.39) &  0.85 (11.12) &  0.79 (11.47) &  0.04 & 1.32 (L1) \cr 
USco J160603.75$-$221930.0 & L2 & 1.15 ( 9.11) &  0.63 (11.49) &  0.88 (10.28) &  0.88 ( 8.45) &  0.04 & 1.31 (L1) \cr 
USco J160956.34$-$222245.5 & dL2 & 1.27 (11.74) &  0.69 ( 9.31) &  0.81 (12.29) &  0.80 (10.90) & -0.03 & 2.43 (T1) \cr 
 \hline
\end{tabular}
\end{table*}

%
%
\begin{table*}
  \caption{Equivalent widths of atomic lines
  for 21 brown dwarfs confirmed as spectroscopic members
  of USco. The L2 field dwarf classified as a non-member,
  USco J160956.34$-$222245.5, is also included. We list 
  the equivalent
  widths of the K {\small{I}} doublets at 1.168/1.177
  and 1.243/1.253 $\mu$m, the Fe {\small{I}} doublet at
  1.189/1.197 $\mu$m, the Al {\small{I}} doublet at
  1.311/1.314 $\mu$m, the K {\small{I}} doublet at 1.517 $\mu$m,
  and Na {\small{I}} doublet at 2.208 $\mu$m (in this order).
  The second column of the table lists the adopted spectral
  type from the direct comparison with previous USco members
  (Table \ref{tab_USco:list_cand}). Objects are ordered by 
  increasing spectral type.
}
  \label{tab_USco:list_atomic}
 \begin{tabular}{c c c c c c c c}
\hline
IAU Name              & Sp  & K {\small{I}} 1.168/1.177 & K {\small{I}} 1.243/1.253 & Fe {\small{I}} & Al {\small{I}} & K {\small{I}} & Na {\small{I}} \cr
\hline
SCH162528.62$-$165850.55   & M8  & 1.8 (0.5) 1.9 (0.5) & 1.2 (0.5) 1.2 (0.5) & 1.2 1.1 & 0.7 & 1.6 & 1.3 \cr 
USco J155419.99$-$213543.1 & M8  & 1.7 (1.0) 4.1 (1.0) & 1.5 (0.5) 2.0 (0.5) & 0.8 0.6 & 0.9 & 2.0 & 5.0 \cr 
USco J160648.18$-$223040.1 & M8  & 3.6 (1.5) 2.6 (1.0) & 1.5 (0.5) 1.6 (0.5) & 0.7 0.7 & 0.6 & 1.8 & 3.3 \cr 
DENIS161103.6$-$242642.9   & M9  & 2.1 (0.5) 3.1 (0.5) & 1.3 (0.5) 1.8 (0.5) & 2.1 1.1 & 0.7 & 0.8 & 2.9 \cr 
USco J160830.49$-$233511.0 & M9  & 1.0 (1.0) 2.3 (0.5) & 1.5 (0.5) 1.1 (0.5) & 1.3 1.0 & 0.7 & 1.5 & 3.5 \cr 
USco J160847.44$-$223547.9 & M9  & 3.7 (1.0) 5.0 (2.0) & 1.9 (0.5) 1.8 (0.5) & 0.8 1.2 & 0.4 & 1.2 & 1.5 \cr 
USco J161047.13$-$223949.4 & M9  & 2.2 (1.0) 1.8 (0.5) & 2.7 (0.5) 3.7 (1.0) & 0.9 1.6 & 0.8 & 2.1 & 3.8 \cr 
USco J154722.82$-$213914.3 & L0  & 5.0 (2.0) --- (---) & 0.7 (1.0) 1.3 (0.5) & --- --- & 1.2 & 0.6 & low \cr 
USco J160606.29$-$233513.3 & L0  & 2.9 (1.0) 7.1 (2.0) & 2.8 (1.0) 2.4 (1.0) & --- --- & 0.4 & 0.8 & 1.1 \cr 
USco J160714.79$-$232101.2 & L0  & 4.8 (2.0) 5.1 (2.0) & 1.4 (1.0) 2.6 (0.5) & 0.9 0.5 & 0.8 & 2.6 & 2.0 \cr 
USco J160737.99$-$224247.0 & L0  & 4.3 (2.0) 1.7 (1.0) & 2.0 (0.5) 3.5 (0.5) & 1.6 0.9 & 0.2 & 3.1 & 3.6 \cr 
USco J160818.43$-$223225.0 & L0  & 2.0 (1.0) 5.2 (1.0) & 2.0 (0.5) 2.3 (0.5) & --- --- & --- & 2.3 & 3.5 \cr 
USco J160828.47$-$231510.4 & L0  & 3.6 (0.5) 1.2 (0.5) & 2.3 (0.5) 1.4 (0.5) & 1.5 1.4 & 0.8 & 2.1 & 2.5 \cr 
USco J160918.69$-$222923.7 & L1  & --- (---) --- (---) & 5.0 (1.0) 5.2 (1.0) & --- --- & 0.4 & 1.3 & 1.6 \cr 
USco J161227.64$-$215640.8 & L0  & 3.6 (1.0) 1.2 (0.5) & 4.8 (1.0) 3.9 (1.0) & --- --- & --- & 3.7 & --- \cr 
USco J161302.32$-$212428.2 & L0  & 1.6 (2.0) 6.8 (2.0) & 2.1 (0.5) 1.9 (0.5) & --- --- & 0.5 & 1.4 & 3.3 \cr 
USco J160723.82$-$221102.0 & L1  & 5.1 (1.0) 4.8 (1.0) & 2.1 (0.5) 1.6 (0.5) & 0.9 3.1 & 0.6 & 2.3 & 3.1 \cr 
USco J160727.82$-$223904.0 & L1  & 2.8 (2.0) 8.5 (1.0) & 2.6 (0.5) --- (---) & 1.5 3.6 & 0.5 & 1.3 & 3.7 \cr 
USco J160843.44$-$224516.0 & L1  & --- (---) --- (---) & 2.2 (2.0) 1.7 (2.0) & --- --- & 0.5 & 1.4 & 2.5 \cr 
USco J161228.95$-$215936.1 & L1  & 3.8 (2.0) 4.5 (1.0) & 2.1 (0.5) 3.2 (1.0) & 1.1 5.6 & 0.3 & 0.8 & 2.5 \cr 
USco J161441.68$-$235105.9 & L1  & 2.1 (0.5) 3.3 (0.5) & 2.9 (1.0) 3.2 (1.0) & 0.6 2.3 & 0.4 & 1.9 & 0.6 \cr 
USco J163919.15$-$253409.9 & L1  & 0.8 (2.0) 6.9 (2.0) & 2.9 (0.5) 2.6 (1.0) & --- --- & 1.2 & 0.6 & 0.9 \cr 
USco J160603.75$-$221930.0 & L2  & 2.7 (1.0) 5.8 (1.0) & 1.3 (1.0) 2.0 (0.5) & 0.8 3.0 & 0.7 & 1.8 & low \cr 
USco J160956.34$-$222245.5 & dL2  & --- (---) --- (---) & --- (---) 7.6 (1.0) & --- --- & 0.9 & 0.8 & 1.0 \cr 
 \hline
\end{tabular}
\end{table*}

%
%
%

\subsection{Membership assessment}
\label{USco:memb_Memb}

The spectral shape and the weakness of gravity-sensitive features
present in the near-infrared spectra of our targets establish the 
membership of 21 out of 23 candidates extracted from the UKIDSS GCS 
by our photometric study \citep{lodieu06,lodieu07a}. Therefore, we 
have added a significant number of spectroscopically-confirmed brown 
dwarfs (2 M8, 3 M9, 16 L0--L2) to the current census of substellar 
members in USco, made of 46 M6, 10 M7, 6 M8, and 2 M9 dwarfs. 
Furthermore, we have extended the current spectral sequence into
the L regime.

Spectroscopy has failed to reveal members with spectral types
as late as those expected from the photometric survey and the
observed infrared colours: $J-K$ colours around 2 mag are typical
for mid- to late-L field dwarfs \citep{knapp04}.
Hence, young L dwarfs should be much redder than their old
counterparts at similar spectral type, implying that our survey 
is insensitive to those mid- to late-L dwarfs and in fact limited 
by the $J$-band rather than the $K$-band completeness limit
($J \simeq$ 18.7 mag compared to $K \simeq$ 17.3 mag).
Indeed, the $K$-band imaging employs microstepping to achieve
better positional accuracy in order to measure proper motions
after the second epoch planned by the GCS\@. Those second epoch 
observations planned in the $K$-band within the framework
of the GCS may be able to resolve this issue by using proper motions
to pick up low-mass brown dwarfs in USco.
Another option for the lack of mid-L dwarfs is that the survey
is actually approaching the bottom of the mass function. However,
it is less probable than the explanation above as less massive
objects have been reported in $\sigma$ Orionis 
\citep[e.g.][]{zapatero07c}.

The success rate of our photometric selection complemented
partly with proper motions (using 2MASS as first epoch) is over 90\%.
This result is better than the level of contamination reported in the
Pleiades and Alpha Per clusters from photometric-based searches
\citep{moraux01,barrado02a}. However, it is consistent with the work
by \citet{adams01a} where photometry and proper motions from photographic
plates and 2MASS were combined to extract a list of members in the
Pleiades. Subsequent spectroscopic follow-up showed a level of
contamination of 13\% down to the completeness limit of
the survey. The low level of contamination of our photometric selection 
looks very promising for future studies of young star forming regions
relatively free of extinction as well as open clusters targeted 
by the UKIDSS GCS\@. We are able to confirm
spectroscopically the low-mass end of the mass function derived
from the sole photometric survey \citep{lodieu07a}.

Among the non-members, we have one reddened early-type star,
USco J161421.44$-$233914.8\@. This object was selected as a 
photometric and proper motion candidate, hence supposedly a foreground 
contaminant.  In addition, we have uncovered a field L2 dwarf, 
USco J160956.34$-$222245.5, by comparison with 2MASS001544.8$+$351603
observed with NIRSPEC at a similar resolution (M03). According to 
the absolute $J$ magnitudes (M$_{J}$) of L2 dwarfs with trigonometric 
parallaxes, we have calculated a distance of 120--140 pc with an 
uncertainty of 30 pc, assuming an error of 0.5 magnitude in M$_{J}$
\citep{vrba04,knapp04}. If this object turns out to be a binary with 
components of equal brightness, it would be more distant by 40\%. 
It is the faintest known field L dwarf of its
subclass by one magnitude\footnote{This statement is based on the
compendium of L and T dwarfs available at 
http://spider.ipac.caltech.edu/staff/davy/ARCHIVE/ and maintained
by C.\ Gelino, D.\ Kirkpatrick, and A.\ Burgasser.
The two coolest L2 dwarfs listed in this webpage are from 
\citet{kirkpatrick00} and \citet{cruz07}. We are not considering
here L dwarfs found in open clusters (e.g.\ Pleiades) and young 
star-forming regions (USco, $\sigma$ Orionis)} and thus the 
furthest discovered to date although upcoming discoveries in the
UKIDSS Large Area Survey will certainly supersede this object in 
terms of distances \citep{kendall07,lodieu07b}. It is also among 
the few field L dwarfs with distances over 100 pc at the time of 
writing \citep{vrba04}. 

%
%
%
\begin{figure}
   \centering
   \includegraphics[width=\linewidth]{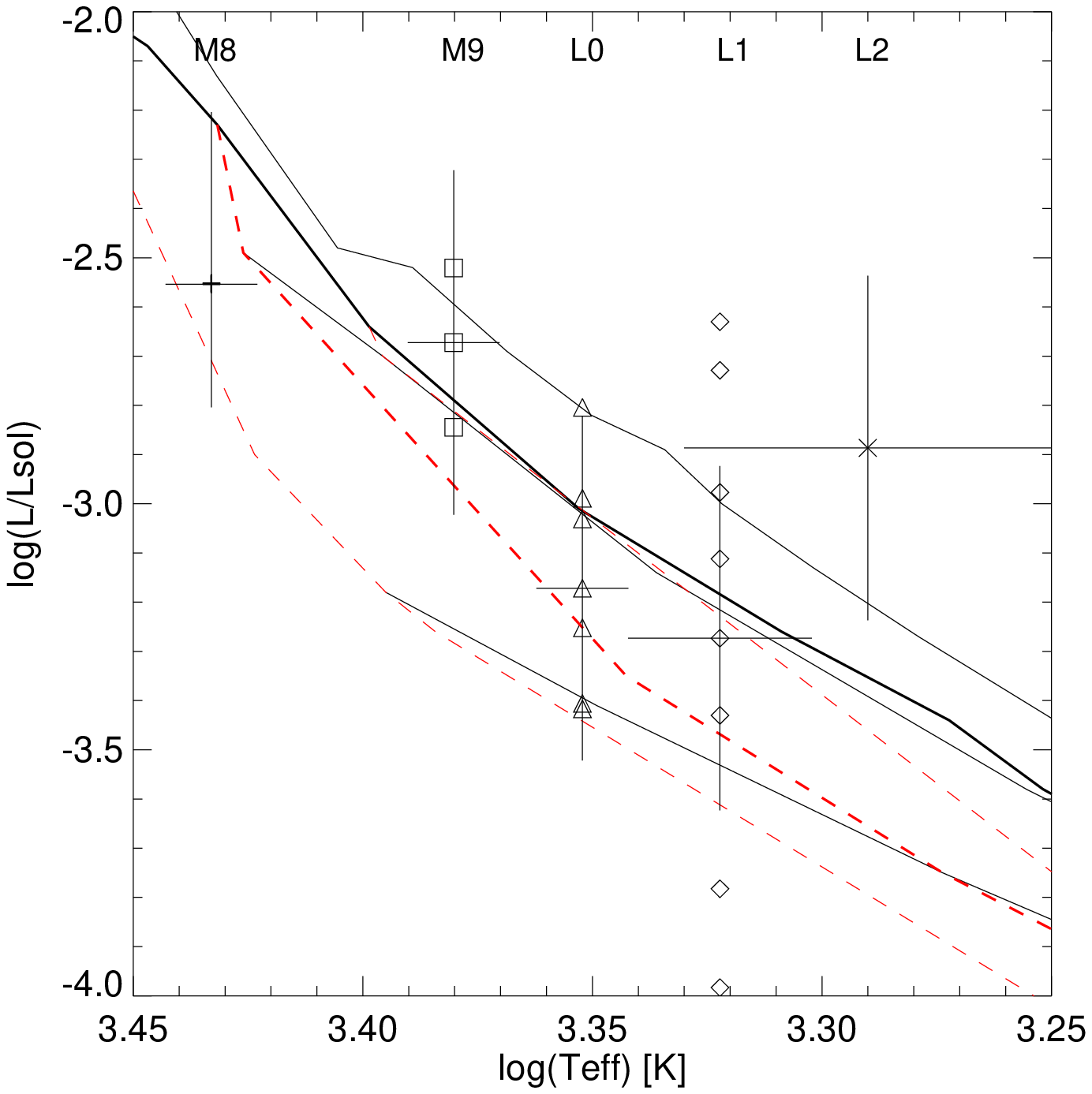}
   \caption{Hertzsprung-Russell diagram (luminosity vs effective
   temperature in logarithmic units) for the new spectroscopic brown 
   dwarfs presented in this paper. Symbols are identical to those in 
   Fig.\ \ref{fig_USco:cmdJKK}: pluses$\equiv$M8, squares$\equiv$M9,
   triangles$\equiv$L0, diamonds$\equiv$L1, and crosses$\equiv$L2\@. 
   Typical error bars on the luminosity (assuming a natural dispersion
   of 0.9 mag; see Sect.\ \ref{USco:discussion}) and on the T$_{\rm eff}$
   are shown for one source of each spectral type to avoid overloading
   the figure. Solid lines are DUSTY isochrones at 1, 5, and 10 Myr
   whereas the dashed lines represent isomasses from the same models 
   for 0.05, 0.03, and 0.02 M$_{\odot}$ \citep{chabrier00c}.
}
   \label{fig_USco:HRdiagram}
\end{figure}
\subsection{Effective temperatures and masses}
\label{USco:memb_Teff}

Combining the luminosities of our targets with their 
T$_{\rm eff}$ deduced from the spectral types, we 
can estimate the masses using theoretical models \citep{chabrier00c}. 
We have applied the T$_{\rm eff}$ versus spectral type relation 
proposed by \citet{luhman99a} and later revised by \citet{luhman03b} 
and specifically designed for young brown dwarfs whose
spectra are intermediate between giants and dwarfs. However, this 
relation is valid only for spectral types earlier than M9, the 
T$_{\rm eff}$ of a M8 and M9 dwarfs being 2710 K and 2400 K, 
respectively. \citet{allers07} extended this relation to young 
L-type dwarfs by subtracting 100 K for each subtype to the
scale established by \citet{luhman03b}.
Recent studies of field dwarfs suggest a difference in T$_{\rm eff}$
between 100 K \citep{golimowski04a} and 200 K \cite{vrba04}.
Subtracting these values to the T$_{\rm eff}$ = 2400 K for M9
yields T$_{\rm eff}$ = 2300--2200 K for L0, 2200--2000 for L1,
and 2100--1800 K for L2\@. The current uncertainty on the
T$_{\rm eff}$ of young early-L dwarfs can thus be as high as
300 K because no L giants are known to create a scale for young
brown dwarfs. At the age \citep[5 Myr;][]{preibisch99} and
distance \citep[145 pc;][]{deZeeuw99} of USco, those intervals
in T$_{\rm eff}$ correspond to masses between 30 and 
8 M$_{\rm Jup}$ \citep{chabrier00c}. The position of our new 
spectroscopic members in the Hertzsprung-Russell diagram is shown 
in Fig.\ \ref{fig_USco:HRdiagram}. We should mention here
that models themselves carry a non negligible uncertainty on their
mass estimates, implying that the mass range quoted above might
be wider.
The source with the latest spectral type (L2) would have a mass
in the 13--8 M$_{\rm Jup}$ range, the lower estimate being 
consistent with the masses derived from the observed $J$ 
magnitude ($J \sim$ 18.5 mag).
To summarise, we assign the following mass ranges 30--20, 20--16.5, 
14--12, and 13--8 M$_{\rm Jup}$ to spectral types of M8, M9, L0, L1, 
and L2, respectively.

%
%
\section{Discussion}
\label{USco:discussion}

Figure \ref{fig_USco:cmdJKK} shows a large dispersion in the
spectral type vs $J$ magnitude relation, particularly around
$J-K$ = 1.4-1.5 mag. In terms of spectral type, both M8 USco members
have $J \sim$ 14.9 mag but if we add SCH1625, the dispersion
would be on the order of 1.3 mag ($J_{\rm 2MASS}$ = 13.67 mag).
Note that difference between 2MASS and WFCAM magnitudes is 
small for late-M dwarfs \citep[0.05 mag or less;][]{hewett06}.
For the three M9 members, we observe a dispersion of 0.8 mag
($J$ = 14.9--15.7 mag; Table \ref{tab_USco:list_cand}). For the L0
dwarfs the dispersion is 1.2 mag ($J$ = 16.0--17.2 mag) and the 
effect is even larger for L1 members with a dispersion
of three magnitudes ($J$ = 15.2--18.5 mag). Even if we remove 
the two reddest sources ($J$ fainter than 18 mag), the dispersion
is on the order of two magnitudes. Although large, this dispersion
is also present among optically-classified USco members: for example
the dispersion for M6 members based on 2MASS \citep{slesnick06}
and DENIS \citep{martin04} photometry span 1.0--1.2 magnitude.
Finally, let's us consider that our sequence is not broadened
by any of the effects described below: assuming that our objects
spread four subclasses (M8--L1; neglecting the single L2 dwarf),
the expected ``natural'' dispersion would lie between 0.65 mag
in $K$ and 0.9 mag in $J$ and represents a significant fraction
of the observed spread in Fig.\ \ref{fig_USco:cmdJKK}.

\begin{itemize}
\item At the age of USco, a dispersion of a few Myr can result in
a large dispersion in magnitudes. \citet{deGeus89} quote nuclear
and kinematic ages between 5 and 8 Myr for Upper Sco whereas
ages for the other two subgroup members of the Scorpius-Centaurus
association, Upper Centaurus Lupus and Lower Centaurus Cruz are
older (10--15 Myr). Similarly, \citet{preibisch99} derived
an age estimate of 5 Myr with little scatter (less than 2 Myr).
The fact that star formation occurred on a short time scale
in Scorpius Centaurus and without a significant age spread
after the explosion of two supernovae \citep{mamajek02} is
summarised in Appendix C of \citet{kraus07a}.
According to theoretical models, the $J$-band magnitudes of a
brown dwarf at 5 and 10 Myr can differ by $\sim$ 0.7 mag. Hence, 
a dispersion of one magnitude is not impossible if the scatter in 
the age is larger than 2 Myr. The scatter in the equivalent width 
measurements of gravity-sensitive doublets 
(Table \ref{tab_USco:list_atomic})
might also be the result of a scatter in age of a few Myr.
Note that we do not consider here the fact that evolutionary models
are less reliable at very young ages ($\sim$ 1 Myr) than at a more
mature stage ($>$100 Myr).
The HR diagram presented in Fig.\ \ref{fig_USco:HRdiagram}
shows a symmetric scatter around the 5 Myr isochrone (thick solid
line), suggesting that age might play a role in the spread observed
in the colour-magnitude but is unlikely to be the main factor.
Two sources appear to be younger and two other much older than
the rest of the members in Fig.\ \ref{fig_USco:HRdiagram}
although error bars are quite large.
\item The distance of USco is known with good accuracy since the
measurements made by Hipparcos 
\citep[145$\pm$2 pc;][]{deBruijne99,deZeeuw99}. 
However, the uncertainty quoted in those works is the uncertainty
on the parallax measurement and does not consider the extent of
the association. The known members from Hipparcos are distributed over 
several degrees across the association, corresponding to an extent
of 15 pc at the distance of 145 pc \citep{preibisch99}.
Larger intervals have been quoted in the literature for USco,
including an uncertainty of 20 pc by \citet{deGeus89} and even
larger range from 80 to 160 pc not inconsistent with the spread
of B-type stars \citep{martin98c}.
A cluster extent of 15 pc would correspond to a spread
of $\sim$0.1 mag in the colour-magnitude diagram
(Fig.\ \ref{fig_USco:cmdJKK}).
\item The presence of reddening might play a role in broadening
the cluster sequence: the typical extinction for USco is less
than A$_{\rm V}$ = 2 mag but translates into $\sim$0.3 mag in
the $J-K$ colour (see Fig.\ \ref{fig_USco:cmdJKK}).
We have applied 2 magnitude of extinction to our M8--L2
spectra and compared them to earlier and later types after
including the effect of reddening: the shape of the spectra
is changed significantly but cannot be reproduced by any other
spectrum. This result suggests that, although reddening could
play a role in broadening the cluster sequence, our new members
are unlikely to be affected by this effect. This conclusion is 
in agreement with the low reddening value derived for SCH1625 from 
its optical spectra \citep[A$_{\rm V}$ = 0.02 mag;][]{slesnick06}.
\item The presence of multiple systems always affects the width
of a cluster sequence. An equal-mass binary will lie 0.75 magnitude
above the single-star sequence, hence contributing to the dispersion
observed in colour-magnitude diagrams.
Although the dispersion observed in the left panel of 
Fig.\ \ref{fig_USco:cmdJKK}
could be explained by the spread in age and the depth of the
association as discussed above, the presence of a gap between the
bluest and reddest sources with $K \sim$ 14.0--15.5 might
reflect the existence of a population of brown dwarf binaries
(unresolved on the seeing-limited images). This fact is best
represented in the ($Y-K$,$Y$) diagram presented in the right panel 
of Fig.\ \ref{fig_USco:cmdJKK}. If the dispersion is indeed mostly
due to binarity, we would derive a multiplicity larger than
50\%. \citet{koehler00} derived a binary frequency of 52$\pm$10\%
over the 0.7--0.13 M$_{\odot}$ from a sample of 118 G5--M5 USco
members and for projected physical separations larger than 21 au.
From a high-resolution survey carried out with the Hubble Space 
Telescope, \citet{kraus05} deduced a low-mass star multiplicity 
of 25$^{+17}_{-9}$\% for separations larger than 7 au but failed to
uncover brown dwarfs below 0.07 M$_{\odot}$. These results point
towards similar properties of young and field brown dwarfs
\citep[low binary fraction, equal-mass systems, 
close separations;][]{burgasser07a}
and is corroborated by other surveys in Cha I \citep{ahmic07}.
However, the contribution of spectroscopic binaries among low-mass
stars and brown dwarfs should not be neglected and can account for 
a significant fraction \citep[$\sim$10--15\%;][]{basri06}.
Finally, we would like to mention that the substellar binary 
fraction in the Pleiades could be as high as 28--44\%
\citep{lodieu07c}, twice larger than estimates from high-resolution
surveys \citep[e.g.][]{bouy06a} but consistent with Monte-Carlo 
simulations by \citet{maxted05} and the radial velocity survey 
by \citet{basri06}.
\item The frequency of disks around low-mass stars and brown dwarfs 
can be as high as 50\% in young star-forming regions, including 
$\sigma$ Orionis \citep{hernandez07,caballero07d}, 
IC\,348 \citep{lada06}, and Chamaeleon \citep{luhman05a}. 
The spectral energy distribution
of several low-mass stars at mid-infrared were indeed best fit 
by disk models in Chamaeleon \citep{natta01}. 
Recently, \citet{carpenter06} found that 19\% of K0--M5
dwarfs in USco show infrared excesses using {\it{Spitzer}}
whereas the majority of higher-mass members ($\geq$ 99\%) seems
devoid of circumstellar disks. During an independent survey,
\citet{jayawardhana03b} found that 50$\pm$25\% (four out of eight 
sources with spectral types later than M5) exhibit $K-L'$ colour 
excesses suggesting the presence of circumstellar disks.
Hence, those results may suggest that at least 20--25\%
of low-mass stars and brown dwarfs possess infrared excesses
longwards of 3.6 $\mu$m.
The resulting dispersion in the colour-magnitude diagrams
can be on the order of several tenths of magnitudes depending
on the colour ($K$ is the most affected band at near-infrared
wavelengths implying an increased dispersion in $J-K$ colour) 
and the mass of the object. For example, \citet{caballero07d}
found a typical dispersion of 0.4 mag in $J-K$ among objects with 
and without disks in $\sigma$ Orionis, a cluster with an age
comparable to USco.
\end{itemize}
%

%
%
\section{Conclusions}
\label{USco:conclusions}

We have presented the results of a near-infrared spectroscopic 
follow-up study aimed at assessing the membership of a sample of brown 
dwarf candidates in the young Upper Sco association. We have confirmed
a total of 21 out of 23 candidates, implying that the luminosity 
function derived form the photometric survey remains unchanged. 
By the same token, the power law index $\alpha$=0.6 derived from
our photometric survey \citep{lodieu07a} remains valid after 
spectroscopic confirmation of over 90\% of the photometric candidates 
below 30 M$_{\rm Jup}$.
The spectral types of the newly confirmed brown dwarfs range from 
M8 to L2, yielding effective temperatures in the 2700--1800 K range
with uncertainties up to 300 K, following
the scale developed by \citet{luhman03b} for young brown dwarfs.
The interval in mass probed by our study spans 30--8 M$_{\rm Jup}$
according to theoretical models and is consistent with the masses
inferred from our photometric survey.
The dispersion observed in the spectral type-magnitude relation
is large, in particular for the latest spectral types. The
reasons for this dispersion might be the consequence of several
effects, including reddening (although unlikely for our sample), 
errors on the age (small contribution), uncertainty on the distance,
binarity, and the presence of disks (affecting at least 20\% of the
sources).

To investigate the binary frequency in USco, several follow-up 
observations are required, including adaptive optics and high-resolution
spectroscopy. Several studies pointed out the possibility of a large
fraction of close and wide low-mass binaries in USco 
\citep{kraus05,bouy06b} compared to other clusters although no
brown dwarf companions were found (in particular in the mass range
discussed here). If binarity is the main effect for the dispersion 
of spectroscopic members observed in colour-magnitude diagrams,
those systems would have separations smaller than th
limits of high-resolution imaging surveys ($\sim$5 au for HST,
7--10 au for adaptive optics). The cause for the difference between
clusters remains however unclear: does it depend on the
environment as proposed by \citet{kraus05} and \citet{bouy06b}
or is it a consequence of the disruption of binaries with age?
The occurrence of disks around low-mass brown dwarfs and planetary-mass 
objects and the influence of disks on the association sequence could
be addressed with a $L'$-band imaging survey
\citep{jayawardhana03b,liu03} or Spitzer follow-up \citep{carpenter06}.
Thus additional photometry at some of these wavelengths for our USco 
brown dwarfs could be made to indicate the presence of disks,
estimate their effect on the $J-K$ colours, and provide some 
indication about the disk properties of our sample.

%
%
\section*{Acknowledgments}

We thank Maria Rosa Zapatero Osorio for early comments on the 
original draft. We also wish to thank the referee for a comprehensive 
and thorough reading of the manuscript.
Many thanks to Stanimir Metchev and Subhanjoy Mohanty for providing
kindly and promptly the spectra of HD\,203030B and 2MASS1207$-$3932B, 
respectively.
This work is based on  observations obtained at the Gemini
Observatory (program GS-2007A-Q-12), which is operated by the
Association of Universities for Research in Astronomy,
Inc., under a cooperative agreement with the NSF on behalf of the
Gemini partnership: the National Science Foundation (United States),
the Particle Physics and Astronomy Research Council (United Kingdom),
the National Research Council (Canada), CONICYT (Chile), the Australian
Research Council (Australia), CNPq (Brazil) and CONICET (Argentina).
This research has made use of the NASA's Astrophysics Data 
System Bibliographic Services (ADS).
This publication has also made use of data products from the
Two Micron All Sky Survey, which is a joint project of the
University of Massachusetts and the Infrared Processing and
Analysis Center/California Institute of Technology, funded by
the National Aeronautics and Space Administration and the National
Science Foundation.
This research has benefitted from the M, L, and T dwarf compendium 
housed at DwarfArchives.org and maintained by Chris Gelino, Davy 
Kirkpatrick, and Adam Burgasser.

%
%
\bibliographystyle{mn2e}
\bibliography{../../AA/mnemonic,../../AA/biblio_old}

%
%

\label{lastpage}

\end{document}